\documentclass[a4paper]{article}

\usepackage[english]{babel}
\usepackage[utf8x]{inputenc}
\usepackage[T1]{fontenc}

\usepackage[a4paper,top=3cm,bottom=2cm,left=3cm,right=3cm,marginparwidth=1.75cm]{geometry}

\usepackage{amsmath}
\usepackage{graphicx}
\usepackage[colorinlistoftodos]{todonotes}
\usepackage[colorlinks=true, allcolors=blue]{hyperref}
\usepackage{stmaryrd}
\usepackage{soul}
\usepackage[normalem]{ulem}
\usepackage{relsize}
\usepackage{lmodern}
\usepackage{slantsc}
\usepackage{graphicx}
\usepackage{amsmath}
\usepackage{amssymb}
\usepackage{amsthm}
\usepackage{amsbsy}
\usepackage{mathrsfs}
\usepackage{varioref}
\usepackage{dsfont}
\usepackage{bm}
\usepackage{color}
\edef\restoreparindent{\parindent=\the\parindent\relax}
\usepackage{parskip}
\restoreparindent
\usepackage[sort, numbers]{natbib}
\usepackage{authblk} 

\newcommand{\sub}[1]{_{\!\mathsmaller{\, #1}}}
\newcommand{\eq}[1]{Eq.~\eqref{#1}}
\newcommand{\fig}[1]{Fig.~\ref{#1}}
\newcommand{\sect}[1]{Sec.~\ref{#1}}

\newcommand{\h}{{\mathcal{H}}}
\newcommand{\hs}{{\mathcal{H}}\sub{\mathcal{S}}}
\newcommand{\he}{{\mathcal{H}}\sub{\mathcal{E}}}
\newcommand{\e}{{\mathcal{E}}}
\newcommand{\s}{{\mathcal{S}}}
\newcommand{\mm}{{\mathcal{M}}}

\newcommand{\vv}{{\mathcal{V}}}

\newcommand{\louv}{{\mathscr{L}}}

\newcommand{\co}{\mathds{C}}
\newcommand{\one}{\mathds{1}}

\newcommand{\tr}{\mathrm{tr}}

\newcommand{\bb}[0]{\begin{eqnarray}}
\newcommand{\be}[0]{\begin{eqnarray}}
\newcommand{\ee}[0]{\end{eqnarray}}
\newcommand{\<}{\langle}
\renewcommand{\>}{\rangle}
\newcommand{\ket}[1]{| #1 \rangle}
\newcommand{\bra}[1]{\langle #1 |}
\newcommand{\moy}[1]{\langle #1\rangle}
\newcommand{\pr}[1]{\Pi[{#1}]}
\newcommand{\prs}[1]{\Pi\sub{\s}[{#1}]}
\newcommand{\pre}[1]{\Pi\sub{\e}[{#1}]}
\renewcommand{\ln}[1]{\mathrm{ln} \left( {#1}\right)}

\newcommand{\ts}{\textsuperscript}

\makeatletter
\DeclareRobustCommand{\cev}[1]{%
  \mathpalette\do@cev{#1}%
}
\newcommand{\do@cev}[2]{%
  \fix@cev{#1}{+}%
  \reflectbox{$\m@th#1\vec{\reflectbox{$\fix@cev{#1}{-}\m@th#1#2\fix@cev{#1}{+}$}}$}%
  \fix@cev{#1}{-}%
}
\newcommand{\fix@cev}[2]{%
  \ifx#1\displaystyle
    \mkern#23mu
  \else
    \ifx#1\textstyle
      \mkern#23mu
    \else
      \ifx#1\scriptstyle
        \mkern#22mu
      \else
        \mkern#22mu
      \fi
    \fi
  \fi
}

\makeatother

\title{Work, heat and entropy production along quantum trajectories}

\author[1]{Cyril Elouard}
\author[2]{M. Hamed Mohammady}
\affil[1]{University of Rochester, Rochester, NY, United States.}
\affil[2]{Lancaster University, Lancaster, United Kingdom.}
\begin{document}
\maketitle

\begin{abstract}
Quantum open systems evolve according to completely positive, trace preserving maps acting on the density operator, which can equivalently be unraveled in term of so-called quantum trajectories. These stochastic sequences of pure states correspond to the actual dynamics of the quantum system during single realizations of an experiment in which the system's environment is monitored. In this chapter, we present an extension of stochastic thermodynamics to the case of open quantum systems, which builds on the analogy between the quantum trajectories and the trajectories in phase space of classical stochastic thermodynamics. We analyze entropy production, work and heat exchanges at the trajectory level, identifying genuinely quantum contributions due to decoherence induced by the environment. We present three examples: the thermalization of a quantum system, the fluorescence of a driven qubit and the continuous monitoring of a qubit's observable.
\end{abstract}

\section{Introduction}

The concept of quantum trajectories was first introduced as a numerical tool to simulate non-unitary quantum evolution \cite{Molmer93}. The underlying idea was to replace the deterministic evolution of a density matrix in a space of dimension $d\sub{\s}^2$ with a stochastic dynamics for a wave function of dimension $d\sub{\s}$, leading to a potentially huge memory gain when simulating large quantum systems. Later on, it was understood that such trajectories actually correspond to the sequences of pure states followed by an open quantum system when its environment is continuously monitored \cite{Wiseman96,Brun02} (see Fig. 1) and the measurement record is accessible. Since then, such a situation has been implemented experimentally in various setups of quantum optics \cite{Nagourney86,Sauter86,Hulet86,Gleyzes07} or superconducting circuits \cite{Murch13,Campagne13,deLange14}. 

Besides, it has long been noticed that an open quantum system (Fig. 1a) can be interpreted as the usual thermodynamic situation: a working substance (the system $\s$) on the one hand driven by an external agent who varies a parameter $\lambda$ in its Hamiltonian (this induces work exchanges), and on the other hand is coupled to a heat bath (this causes heat dissipation) \cite{Alicki79}. This canonical situation was extensively studied in the classical case, first restricting the working substance to near-equilibrium states and more recently, owing to the statistical thermodynamics paradigm  \cite{Seifert08,Sekimoto10}, for any transformation potentially bringing the system $\s$ far from equilibrium. 

In the formalism of classical statistical thermodynamics, the classical system follows a stochastic trajectory in its phase space at each single realization of the studied thermodynamic transformation. The randomness in this picture originates from the nature of the actions of the environment on the system $\s$ that appear in practice to be uncontrollable and unpredictable. Thermodynamic quantities like work, dissipated heat, and  entropy production become stochastic variables defined for a single trajectory. These quantities satisfy fluctuation theorems \cite{Jarzynski97,Crooks99,Seifert05} which constrain their out-of-equilibrium fluctuations beyond the second law of thermodynamics. 

When the environment is  monitored, the quantum trajectories can be seen as quantum analogues of the stochastic trajectories in phase space, enabling to build a quantum version of the stochastic thermodynamics formalism \cite{Hekking13,Alonso16,Manzano-fluctuation-quantum-maps,Alexia-measurement-thermodynamics,Gherardini17}. However, these quantum trajectories feature two crucial differences with respect to the trajectories of classical stochastic thermodynamics: First, they correspond to sequences of states in the system's Hilbert's space $\hs$ instead of phase space. They therefore allow to study the thermodynamic consequences of quantum specificities such as coherences and entanglement. Second, the stochasticity of the system's dynamics is a direct consequence of the randomness of quantum measurements, and is not predicated on the observer's ignorance as to the thermal environment's precise evolution. As a consequence, quantum stochastic thermodynamics can also describe situations where there is no thermal reservoir, but the system is coupled to a measuring apparatus which introduces randomness and, therefore, irreversibility in the system's dynamics \cite{Alexia-measurement-thermodynamics,Dressel17,Manikandan18}. Such formalism therefore opens an avenue to describe the thermodynamics of quantum measurements.

In this chapter, we review the formalism of quantum stochastic thermodynamics and highlight some of the genuinely quantum features it opens to study owing to fundamental examples. The chapter is divided as follows: in Section \ref{s:Qtraj}, we introduce the concept of quantum trajectories and unraveling, first for any quantum channel, and then in the special case of a quantum open system obeying a Lindblad equation. In Section \ref{s:Entr}, we define the entropy production at the level of a quantum trajectory and show it obeys an Integral fluctuation theorem as in the classical case. In Section \ref{s:FirstLaw}, we analyze the energetics along a single trajectory and stress the contribution of coherence erasure induced by the quantum measurements. In Sections \ref{s:Ex1}-\ref{s:Ex3} we analyze three fundamental examples: the thermalization of a quantum system, the photo-counting of a qubit's fluorescence and the continuous measurement of qubit's obervable.

\begin{figure}[!htb]
\begin{center}
\includegraphics[width = 0.9\linewidth]{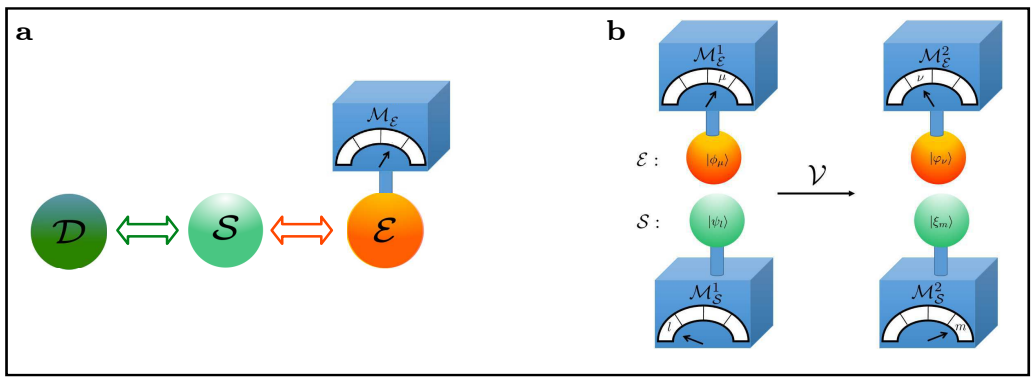}
\end{center}
\caption{\textbf{a}: Typical scenery of quantum open systems formalism: a system $\s$ is coupled to an environment $\e$ and potentially driven by an external agent $\cal D$. In the quantum trajectory paradigm, $\e$ is monitored by a measuring apparatus ${\cal M}_\e$. \textbf{b}: The joint quantum trajectory $\Gamma := ((l,m),(\mu,\nu))$ observed when the system $\s$ and environment $\e$ are first measured by the detectors $\mm\sub{\s}^1$ and $\mm_\e^1$, respectively, after which they evolve by the unitary channel $\vv$. Finally, they are measured by the detectors $\mm\sub{\s}^2$ and $\mm_\e^2$ (this time measuring a different observable). }
\label{fig:simple-traj}
\end{figure}

\section{Quantum trajectories} 
\label{s:Qtraj}

\subsection{Two-point quantum trajectories of open systems}\label{s:Qtraj-open-system}

Consider a  quantum system $\s$, with  a  Hilbert space $\hs \simeq \co^{d\sub{\s}}$, that is initially prepared in the state $\rho\sub{\s} = \sum_{l=1}^{d\sub{\s}} p_l \prs{\psi_l}$. Here $\prs{\psi} \equiv |\psi\>\<\psi|$ is a projection on the vector $\ket{\psi} \in \hs$ and $\{\ket{\psi_l}\}_l$ is an orthonormal basis that spans $\hs$. If the system is closed, its time evolution will be described by the unitary quantum channel $\vv( \rho\sub{\s}) :=  V \rho\sub{\s} V^\dagger = \sum_{m=1}^{d\sub{\s}} p_m \prs{\xi_m}$, with $\{\ket{\xi_m}\}_m$ another  orthonormal basis that spans $\hs$. Now consider we projectively measure the system with respect to the observables $\{\prs{\psi_l}\}_{l=1}^{d\sub{\s}}$ and $\{\prs{\xi_m}\}_{m=1}^{d\sub{\s}}$ prior and posterior to the time evolution, respectively. The evolution of the average (unconditioned) state will still be given by $\rho\sub{\s} \mapsto V \rho\sub{\s} V^\dagger$; this is because the observables commute with these states. Consequently, we can decompose the total evolution of the system into a collection of measurement sequences, or ``two-point pure-state trajectories'', $\Gamma := (l,m) \equiv \ket{\psi_l} \mapsto \ket{\xi_m}$. Each trajectory $\Gamma$ corresponds to a possible evolution of the system during a single realization of the process consisting of the first measurement, the evolution and the second measurement, the one characterized by the sequence of measurement outcomes $(l,m)$. The probability of the trajectory $\Gamma$ will be given by the Born rule as
\begin{align}\label{eq:prob-forward-trajectory-unitary}
P(\Gamma) &= \tr[\prs{\xi_m} \vv( \prs{\psi_l}\rho\sub{\s} \prs{\psi_l})], \nonumber \\&= \tr[\prs{\xi_m} \vv(\prs{\psi_l}) ] \, \tr[\prs{\psi_l}\rho\sub{\s}] \nonumber \\
& = \tr[\prs{\xi_m} \vv( \prs{\psi_l})] \, p_l .
\end{align}

Now consider the case where $\s$ is an open system \cite{breuer}, such that it interacts with an environment $\e$ with Hilbert space $\he$. If the compound system $\s+\e$ is initially prepared in the product state $\rho\sub{\s}\otimes \rho\sub{\e}$, and then evolves by a joint unitary channel $\vv: \rho\sub{\s}\otimes \rho\sub{\e}\mapsto V(\rho\sub{\s}\otimes \rho\sub{\e})V^\dagger$, the evolution of $\s$ alone will be described by the quantum channel $\Phi$ \cite{Heinosaari}, defined as
\begin{align}\label{eq:quantum channel}
\Phi(\rho\sub{\s}) := \tr\sub{\e}[V(\rho\sub{\s} \otimes \rho\sub{\e})V^\dagger].
\end{align}
Here  $\tr\sub{\e}$ is the partial trace over $\he$. The triple $(\h\sub{\e}, \rho\sub{\e},V)$ is referred to as the Stinespring dilation of $\Phi$ \cite{Stinespring}. Whilst a quantum channel can arise from infinitely many such dilations, a given choice of environment and joint evolution describes a unique channel.  

We now wish to determine the two-point measurement scheme that is compatible with the \emph{reduced}  dynamics of $\s$ as defined by the quantum channel $\Phi$. To this end, we must projectively measure $\s$, at the start and end of its evolution, with observables that do not disturb its average state. The initial observable is, as before, $\{\prs{\psi_l}\}_l$, and given that the final state of $\s$ is written as $\Phi(\rho\sub{\s})=\sum_m p_m' \prs{\xi_m}$, the final observable is $\{\prs{\xi_m}\}_m$.  The resulting sequence of measurements outcomes will therefore define a trajectory of $\s$, denoted $\Gamma_\s :=(l,m) \equiv \ket{\psi_l}\mapsto \ket{\xi_m}$, which will occur with the probability
\begin{align}\label{eq:prob-forward-trajectory-Channel-total}
P(\Gamma_\s) = \tr[\prs{\xi_m} \Phi(\prs{\psi_l})] \,  p_l.
\end{align}
Let us assume that we are also capable of independently measuring the environment. Given that the initial and final state of the environment is  $\rho\sub{\e} = \sum_\mu q_\mu \pre{\phi_\mu}$ and $\rho\sub{\e}' = \sum_\nu q_\nu' \pre{\varphi_\nu}$, respectively, by measuring $\e$ with the observable $\{\pre{\phi_\mu}\}_\mu$, respectively $\{\pre{\varphi_\nu}\}_\nu$, at the start and end of the evolution, we may  construct the trajectories $\Gamma_\e := (\mu, \nu) \equiv \ket{\phi_\mu} \mapsto \ket{\varphi_\nu}$. The \emph{joint trajectory} for both system and environment, therefore, is $\Gamma := (\Gamma_\s, \Gamma_\e) = ((l,m),(\mu,\nu))$, which occurs with the probability
\begin{align}\label{eq:prob-forward-trajectory}
P(\Gamma) &= \tr[(\prs{\xi_m}\otimes \pre{\varphi_\nu})V(\prs{\psi_l}\otimes \pre{\phi_\mu})V^\dagger] \, \tr[\prs{\psi_l}\rho\sub{\s}]\tr[\pre{\phi_\mu}\rho\sub{\e}], \nonumber \\
&= \tr[(\prs{\xi_m}\otimes \pre{\varphi_\nu})V(\prs{\psi_l}\otimes \pre{\phi_\mu})V^\dagger] \, p_l q_\mu.
\end{align}
This scheme is illustrated in \fig{fig:simple-traj}. Note that while the evolution of $\s$, respectively $\e$, can be seen as a stochastic sampling from the trajectories $\Gamma_\s$ and $\Gamma_\e$, the evolution of the compound system $\s+\e$ is not given by a stochastic sampling from the joint trajectories $\Gamma$. This is because, generally, the compound system will be correlated at the end of the evolution, and the measurements will disturb the state.

In order to express \eq{eq:prob-forward-trajectory} analogously to \eq{eq:prob-forward-trajectory-Channel-total}, i.e., in terms of dynamical processes on the system alone, we must first obtain the ``Kraus'' decomposition of the channel $\Phi$ as  $\Phi(\rho\sub{\s}) = \sum_{\mu, \nu} \Phi_{\mu, \nu} (\rho\sub{\s})$, where
\bb\label{eq:Phimunu}
 \Phi_{\mu, \nu} (\rho\sub{\s}) = M_{\mu, \nu} \rho\sub{\s} M_{\mu, \nu}^\dagger,
 \ee
such that  
\bb\label{eq:Mmunu}
M_{\mu, \nu}:= \sqrt{q_\mu} \<\varphi_\nu| V |\phi_\mu\>
\ee
are the Kraus operators \cite{Kraus} determined by the environment trajectory $\Gamma_\e$. Consequently, \eq{eq:prob-forward-trajectory} can be expressed  as
\begin{align}\label{eq:prob-forward-trajectory-Channel}
P(\Gamma) = \tr[\prs{\xi_m} \Phi_{\mu,\nu}(\prs{\psi_l})] \, p_l.
\end{align}

We note that $\{E(\nu):=\sum_{\mu} M_{\mu,\nu}^\dagger M_{\mu,\nu}\}_{\nu}$ constitutes a generalized positive operator valued measure (POVM) on the system, with $\nu$ denoting the measurement outcomes \cite{Busch-operational,Busch-measurement-2,WisemanBook}. The effect operators $E(\nu)$ will in general not be projection operators.

\subsection{Quantum trajectories and continuous measurement}

In this section we generalize the preceding considerations to the case where the environment is subject to an arbitrarily long sequence of measurements, rather than just at the beginning and end of the transformation under study.  To be sure, measuring the environment at the middle of the joint evolution will in general destroy the quantum correlations between the system and the environment states built by the unitary evolution. Consequently, unlike the two-point measurement scheme discussed above, we will not be able to claim that our measurements merely track the evolution of the subsystems in general. However, if the correlations between the environment the system vanish on a time-scale much shorter than the time between two measurements on the environment, as it is the case for a Markovian environment, the repeated measurements will as before preserve the average evolution of the system.

To model this situation, we discretize the evolution time, introducing the times $t_k = t_i + k\,\Delta t$ ($t_i \equiv t_0$, $t_f \equiv t_K$) at which the measurements are performed on the environment. More precisely, we assume that a the start $t_{k-1}$ of each time interval $[t_{k-1},t_k]$, the system and environment are in a product state $\rho_\s(t_{k-1})\otimes\rho_\e$. A first measurement is performed on the environment at time $t_{k-1}^+$, yielding outcome $\mu_k$. Then the system and environment evolve unitarily during $\Delta t$ until time $t_{k}^-$ where a second measurement is performed, yielding outcome $\mu_k$. This in turn induces the quantum channel $\Phi(t_{k}, t_{k+1})$ on the system. Consistently with the Markovian assumption, the system and environment are assumed to be in a product state $\rho_\s(t_{k-1})\otimes\rho_\e$ at the beginning of the next evolution interval, i.e. time $t_{k}$, such that the evolution. One often model such process by allowing the system to interact with a new copy of the environment every $\Delta t$, enforcing the Markovian character of the evolution.

The full evolution between $t_i$ and $t_f$ is given by the composition of these channels, i.e.,  $\Phi(t_i, t_f) = \Phi(t_{K-1},t_K)\circ \dots \circ \Phi(t_1,t_2)\circ \Phi(t_0,t_1)$.
A single realization of the process is now determined by a measurement record $\vec{\alpha}= \{\alpha_k\}_{k=1}^K$, where for each $k$,  $\alpha_k = (\mu_k,\nu_k)$ compiles the outcomes obtained at the beginning and at the end of the interval $(t_{k-1},t_k)$. The Kraus decomposition of each map $\Phi(t_{k-1},t_{k})$ associated with the  measurement outcomes $\alpha_k$ obtained between $t_{k-1}$ and $t_{k}$ can be introduced similarly as in Eq.\eqref{eq:Phimunu}: $\Phi(t_{k-1},t_{k}) = \sum_{\alpha_k}\Phi_{\alpha_k}(t_{k-1},t_{k})$.
Let the average state of the system at time $t_0$ and $t_K$ be $\rho\sub{\s}(t_0) = \sum_l p_l \prs{\psi_l}$ and $\rho\sub{\s}(t_K) = \sum_m p_m' \prs{\xi_m}$, respectively.  As such, we obtain the trajectories $\Gamma := ((l,m),\vec{\alpha})$, with probabilities
\begin{align}\label{eq:ptraj}
P(\Gamma) = \tr[\prs{\xi_m}(\Phi_{\alpha_K}(t_{K-1},t_{K})\circ \dots \circ \Phi_{\alpha_1}(t_0,t_1))( \prs{\psi_l})] \, p_l.
 \end{align}

Since each $\Phi_{\alpha_k}(t_{k-1},t_k)$ is described by a single Kraus operator $M_{\alpha_k}$, we may view the evolution of the system along the time sequence $\{t_k\}_{k=0}^K$ as a trajectory of pure states 
\begin{align}\label{eq:Lindblad-system-pure-state-traj}
\ket{\Psi_\Gamma(t_0)} \mapsto \ket{\Psi_\Gamma(t_1)} \mapsto \dots \mapsto \ket{\Psi_\Gamma(t_K)},
\end{align}
 such that
\begin{align}\label{Psik_traj}
\ket{\Psi_\Gamma(t_{k})} = \dfrac{M_{\alpha_{k}} \ket{\Psi_\Gamma(t_{k-1})}}{\| M_{\alpha_k}\ket{\Psi_\Gamma(t_{k-1})}\|^2},
\end{align}
where $\ket{\Psi_\Gamma(t_0)} = \ket{\psi_l}$ and $\ket{\Psi_\Gamma(t_K)} = \ket{\xi_m}$. Generally, the time $\Delta t$ is chosen to be much shorter than the characteristic evolution time of the system, meaning that each $M_{\alpha_k}$ is very close to the identity operator. As such, the states $\ket{\Psi_\Gamma(t_k)}$ along this trajectory vary slowly, and can be in a superposition of states from the basis in which the environment induces decoherence.

A great interest of such a time-resolved measurement scheme is that it generally involves the monitoring of the environment subspace interacting with the system during the interval $\Delta t$, rather than the complete environment, which is amenable to experimental implementation. Conversely, performing projective measurements on the total Hilbert space of the environment -- which is generally large -- is usually impractical.  Examples of such continuous measurement schemes include monitoring of the fluorescence emitted by a superconducting qubit in a transmission line \cite{Murch13}, and the readout of Rydberg atoms after they interact sequentially with a cavity field \cite{Gleyzes07}. In these setups, it is only required that we measure the field in the transmission line and a single atom every $\Delta t$, respectively, to have enough information to reconstruct the system's trajectory.

We finally note that we have considered rather ideal measurements. The present formalism can be extended to account for various practical limitations. An extension to measurements associated to non rank-1 projectors in $\hs$ is presented in the Appendix A to this chapter. Other common limitations includes imperfect detection, which is can be modeled by an average over a subset of the measurement record (see e.g. \cite{WisemanBook}, section 4.8 and \cite{Steck06}, section VII).

\subsection{Quantum trajectories from a Lindblad equation}
\subsubsection{Kraus decomposition of the Lindblad equation}
For a quantum system in contact with a Markovian environment at equilibrium, the dynamics may often be described by a Lindblad equation \cite{breuer} $\partial_t\rho\sub{\s} = \louv_\lambda(\rho\sub{\s}(t))$ where the Liouvillian $\louv_\lambda$ fulfills:
\begin{align}\label{eq:Lindblad}
\louv_\lambda(\rho\sub{\s}) = i [\rho\sub{\s}, H\sub{\s}(\lambda)]\sub{-} + \sum_{j=1}^J {\cal D}_j^\lambda(\rho\sub{\s}).
\end{align}

\noindent Here $H\sub{\s}(\lambda)$ is the system's Hamiltonian which depends on a parameter $\lambda$ controlled by an external agent. This parameter is in general varied during a thermodynamical transformation, and takes the values $\{\lambda_t\}_{t_i\leqslant t\leqslant t_f}$. ${\cal D}_j^\lambda$ is a superoperator satisfying:
\be\label{eq:Dj}
{\cal D}_j^\lambda(\rho\sub{\s}) := L_j(\lambda) \rho\sub{\s} L_j^\dagger(\lambda) - \frac{1}{2}[\rho\sub{\s}, L_j^\dagger(\lambda) L_j(\lambda)]\sub{+}.
\ee
\noindent Here $[\cdot,\cdot]\sub{-}$ and $[\cdot,\cdot]\sub{+}$ are respectively the commutator and anti-commutator and $L_j(\lambda)$ are the so-called Lindblad (or jump) operators. The Liouville super-operator given in \eq{eq:Lindblad} generates the quantum channel $\Phi(t_i,t_f): \rho\sub{\s} \mapsto  \mathbb{T}e^{\int_{t_i}^{t_f} dt\louv_{\lambda_t}}(\rho\sub{\s})$, with $\mathbb{T}$ the time-ordering operator. $\Phi(t_i,t_f)$ is a composition of infinitesimal quantum channels $\Phi^{\lambda_k}(t_{k-1},t_{k}) = \one\sub{\s}+dt\louv_{\lambda_k}$, with $\lambda_k \equiv \lambda_{t_k}$.

The exact form of the Liouvillian in \eq{eq:Lindblad} can be derived from a given microscopic model of the system and the environment by averaging the  exact unitary dynamics of the joint system over a time that is large with respect to the correlation time of the environment, but short with respect to the relaxation time of the system, and tracing over the environment \cite{breuer,cohen}. Alternatively, when the microscopic description of the environment is unknown, the Lindblad equation associated with a set of $A = d\sub{\s}^2-1$ Lindblad operators, $d\sub{\s}$ being the dimension of $\hs$, can be used phenomenologically as the most general infinitesimal quantum channel capturing the dynamics of a system in contact with a Markovian environment \cite{Lindblad76}.

The quantum trajectories are obtained from a Kraus decomposition of $\sum_{\alpha_k}\Phi_{\alpha_k}^{\lambda_k}(t_{k-1},t_{k})$ associated with measurement outcomes $\{\alpha_k\}$. The environment, if Markovian, is assumed to be respectively in the states $\rho\sub{\e} = \sum_\mu q_\mu \pre{\phi_\mu}$ and $\rho\sub{\e}' = \sum_\mu q_\mu' \pre{\phi_\mu}$ at the infinitesimal time-steps $t_{k-1}$ and $t_k$. The Markovian environment is affected very slightly by its interaction with the system, quantified by the fact that the relative entropy $D[\rho\sub{\e}'\| \rho\sub{\e}]:= \tr[\rho\sub{\e}'(\ln{\rho\sub{\e}'} - \ln{\rho\sub{\e}})]$ is vanishingly small. As such, if the environment were projectively measured with respect to the same observable $\{ \pre{\phi_\mu}\}_\mu$ before and after the infinitesimal interaction time, we would have $\alpha_k = (\mu_k, \nu_k)$, analogously to our earlier discussion in \sect{s:Qtraj-open-system}.   However, owing to the Markovian character of the environment, one can measure the environment in many different ways without affecting the system's master equation.  
The choice of a particular measurement scheme on the environment determines the so-called ``unravelling'' of the Lindblad master equation. As there are infinitely many ways of measuring the environment, there are infinitely many such unravellings. In the following sections, we will focus on two measurement schemes that are particularly relevant for situations of interest in thermodynamics, and have been implemented experimentally: (i) the Quantum Jump (QJ) and (ii) the Quantum State Diffusion (QSD) unravellings.

\subsubsection{Quantum Jumps unraveling}\label{s:QJ}

The QJ unraveling corresponds to measurement of the environment with $J+1$ possible outcomes. 
The first $J$ outcomes $\{\alpha_j\}_{j= 1}^J$ have an infinitesimal probability of occurrence, of order $dt$, and are associated with a strong effect on the system called a ``quantum jump''\cite{Haroche,Molmer93,Kist99,Brun00,WisemanBook}. Such quantum jumps are captured by the Kraus operators:

\begin{align}
M_{\alpha_j}(\lambda) &= \sqrt{dt} L_j(\lambda).\label{eq:QJ_Mj}
\end{align}

\noindent The last outcome $\alpha_0$, called the ``no-jump'' outcome, has a probability of order $1$ and is associated with an infinitesimal evolution characterized by the Kraus operator:

\be
M_{\alpha_0}(\lambda) &= \one\sub{\s} -dt \left(i H\sub{\s}(\lambda) + \frac{1}{2} \sum_{j=1}^J L_j^\dagger(\lambda) L_j(\lambda)  \right).\label{eq:QJ_M0}
\ee

\noindent The "no-jump" evolution can also be seen as induced by the non-Hermitian effective Hamiltonian:

\begin{align}
H\sub{\mathrm{eff}}(\lambda_t):= H\sub{\s}(\lambda_t)  -\frac{i}{2} \sum_{j=1}^J L_j^\dagger(\lambda_t) L_j(\lambda_t).
\end{align}

Such an unraveling typically arises when the measurement on the environment corresponds to counting of the excitations exchanged between the environment and the system, e.g. photon-counting. The $L_j$ operators are then annihilation and creation operators for these excitations. Note that in most cases, only the variation of the environment's excitation number can be accessed (rather than the absolute initial and final number of excitations in the environment), such that a proper derivation of the Kraus operators of the QJ unraveling requires to slightly extend the paradigm introduced in Section \ref{s:Qtraj-open-system} (see Appendix A).

\subsubsection{Quantum state diffusion unraveling}

A QSD unraveling is obtained when there is a continuous set of possible outcomes $\alpha_k$ for the measurement performed on the environment, at time $t_k$ \cite{Gardiner,WisemanBook,Wiseman96,Steck06}. As a result the system's evolution is determined by an infinite number of Kraus operators. The evolution of the system takes the form of a stochastic differential equation in term of the record $\alpha_k$. When using Ito's convention for stochastic calculus, one can in general cast the Kraus operators under the form:
\be\label{eq:MQSD}
M_{\alpha_k}(\lambda_k) = \left(\prod_{j=1}^J p_0(dw_j)\right)^{1/2}\left(\one\sub{\s} -dt \left(i H\sub{\s}(\lambda_k) + \frac{1}{2} \sum_{j=1}^J L_j^\dagger(\lambda_k) L_j(\lambda_k)  \right) +\sum_{j=1}^J dw_j(\alpha_k)L_j(\lambda_k)\right),
\ee
where the $dw_j(\alpha_k)$, $j\in\llbracket 1,J\rrbracket$ are Wiener increments, i.e. complex Gaussian stochastic variables of zero expectation value and fulfilling Ito's rule $dw_j(\alpha_k)^2 = dt$. We denote $p_0(u) = \exp(-u^2/2 dt)/\sqrt{2\pi dt}$ the probability density fulfilled by the Wiener increments.

A typical situation leading to a QSD unraveling is when the environment of the system is actually a meter such that the measurement outcome at time $t_k$ corresponds to a weak measurement of an observable $X$ on the system. In such situation, 
there is only one Lindblad operator $\sqrt{\gamma_\text{mes}}X$ associated with the Wiener increment $dw(\alpha_k) = 2\sqrt{\gamma_\text{mes}}dt(\alpha_k - \moy{X})$ \cite{Steck06}. Other common situations leading to QSD unraveling are the homodyning and heterodyning measurements of the fluorescence of an atom \cite{Wiseman96}.

\section{Entropy production and Integral fluctuation theorem} 
\label{s:Entr}

\subsection{Time-reversed trajectories, stochastic entropy production, and the second law}\label{s:reversed-traj-entropy-prod}
Each trajectory $\Gamma$ has associated with it a stochastic entropy production $\Delta_i s[\Gamma]$, which quantifies the degree of irreversibility in the transformation. 
In concordance with the formalism of classical stochastic thermodynamics \cite{Sekimoto10,Seifert05}, this quantity is obtained by comparing the probability of a trajectory $\Gamma$ to that of its time-reversed counterpart $\tilde \Gamma$, seen as a trajectory generated by the time-reversed thermodynamic transformation.  Namely, we define:
\begin{align}\label{eq:ent-prod}
\Delta_i s[\Gamma] := \ln{\frac{P(\Gamma)}{\tilde P(\tilde \Gamma)}}.
\end{align}
Consequently, the more probable the forward trajectory occurs, in comparison with the time-reversed trajectory, the more entropy is produced. We show below that this definition is consistent with the definition of average entropy production as the total increase in the von Neumann entropy of  system and environment, evaluated with respect to their reduced states. For further justification of using \eq{eq:ent-prod} as the definition of stochastic entropy production, we refer to \cite{CH15}.

Several approaches have been explored to define the time-reversed trajectories and their probabilities. 
The approach introduced in \cite{Crooks08} and used in \cite{Manzano15,Alexia-measurement-thermodynamics,Gherardini17} exploits a fixed point of the quantum map to define the reversed Kraus operators, ensuring that no entropy is produced on average when the system is in the fixed point. This approach has allowed pioneer analyses of the entropy production at the level of single quantum trajectories. However, this approach may break down when the system is driven, in particular for quantum maps that do not have a fixed point \cite{Barra17}. The approach introduced in \cite{Dressel17,Manikandan18} exploits the inverse of the Kraus operator, and is therefore well suited for weak measurements, while it cannot handle rank-1 Kraus operators as those involved in the QJ unraveling. Here, in line with \cite{Horowitz12,Barra17,Manzano17}, we exploit the fact that when both the system and environment are measured with rank-1 projective measurements, the Kraus operator generating the time-reversed sequence of states of the environment is uniquely defined. This allows to evaluate the entropy production associated to a large variety of quantum processes as shown by the examples considered in this chapter.

Recall that a general trajectory $\Gamma = ((l,m),\vec\alpha)$, where $\vec \alpha := (\alpha_1,\dots,\alpha_K)$, is defined by two measurements at the start and end of the process on the system, and a sequence of quantum maps $\{\Phi_{\alpha_k}(t_{k-1}, t_{k})\}_{k=1}^K$ that describe the stochastic transformation of the system as time moves forward from $t_{k-1}$ to $t_{k}$.  The time-reversed trajectory $\tilde \Gamma = ((m,l),\cev\alpha)$, where $\cev{\alpha} := (\alpha_K,\dots,\alpha_1)$, is similarly defined with respect to a sequence of  time-reversed quantum maps $\{\tilde{\Phi}_{\alpha_k}(t_k, t_{k-1})\}_{k=K}^1$ that describe the stochastic transformation of the system as time moves backwards from $t_k$ to $t_{k-1}$.  Given that the initial state of the system during the time-reversed process is the average state it occupies at the end of the forward process, namely $\rho\sub{\s}(t_K) := \Phi(\rho\sub{\s}) = \sum_m p_m' \prs{\xi_m}$, the probability for the time-reversed trajectory is obtained analogously to \eq{eq:ptraj} as
\bb\label{eq:time-reversed-probability-general}
\tilde P(\tilde \Gamma) = \tr[\Pi\sub{\s}[\psi_l]( \tilde{\Phi}_{\alpha_1}(t_{1},t_0)\circ\dots\circ\tilde{\Phi}_{\alpha_K}(t_{K},t_{K-1}))(\Pi\sub{\s}[\xi_m])]p'_m.
\ee

Let us assume for simplicity that each quantum channel $\Phi(t_{k-1}, t_k)$ is described by the same unitary $V$, and initial environment state $\rho\sub{\e} = \sum_\mu q_\mu \pre{\phi_\mu}$. This would describe the case where the environment is Markovian. 
As such, \eq{eq:ptraj} can be written as
\begin{align}\label{eq:forward-traj-prob-kraus}
P(\Gamma) = \tr[\prs{\xi_m} M_{\alpha_K}\dots M_{\alpha_1} \prs{\psi_l} M_{\alpha_1}^\dagger \dots M_{\alpha_K}^\dagger] \,  p_l,
\end{align}
where $M_{\alpha_k} = \sqrt{q_{\mu_k}} \<\varphi_{\nu_k}|V|\phi_{\mu_k}\>$. Here we recall that $q_{\mu_k}$ is the probability that the environment at time $t_{k-1}$ occupies the state $\ket{\phi_{\mu_k}}$.

We now introduce the time-reversal paradigm applied in this chapter. To construct the infinitesimal maps generating the time-reversed trajectory, we assume that the initial state of the environment, at time $t_k$, is the average state it occupies, during the forward process at the end of interval $(t_{k-1}, t_k)$, i.e., $\rho\sub{\e}(t_k) = \tr\sub{\s} [V (\rho\sub{\s}(t_{k-1}) \otimes \rho\sub{\e}) V^\dagger] = \sum_{\nu_k} q_{\nu_k}'\pre{\varphi_{\nu_k}}$. Consequently, we may define the $k$\ts{th} time reversed quantum channel as 
\begin{align}\label{eq:reverse quantum channel}
\tilde\Phi(t_k, t_{k-1})(\sigma\sub{\s}) &:= \tr\sub{\e}[V^\dagger(\sigma\sub{\s} \otimes \rho\sub{\e}(t_k))V], \nonumber \\
&= \sum_{\alpha_k}\tilde \Phi_{\alpha_k}(t_k, t_{k-1})(\sigma\sub{\s}),
\end{align}
where we note that the unitary interaction $V$ has been subjected to the time-reversal operation, thus transforming it to $V^\dagger$ \cite{Manzano17}. Each $\tilde \Phi_{\alpha_k}(t_k, t_{k-1})$ is described by the Kraus operator
\bb\label{eq:Mtilde}
\tilde M_{\alpha_k}:= \sqrt{q_{\nu_k}'} \<\phi_{\mu_k}| V^\dagger |\varphi_{\nu_k}\>\equiv \sqrt{\frac{q_{\nu_k}'}{q_{\mu_k}}}M_{\alpha_{k}}^\dagger.  
\ee
Therefore, we may write \eq{eq:time-reversed-probability-general} as
\begin{align}\label{eq:reverse-traj-prob-kraus}
\tilde P (\tilde \Gamma) &= \tr[\prs{\psi_l} \tilde M_{\alpha_1}\dots \tilde M_{\alpha_K} \prs{\xi_m} \tilde M_{\alpha_K}^\dagger \dots \tilde M_{\alpha_1}^\dagger] \,  p_m', \nonumber \\
&= \frac{q_{\nu_1}' \dots q_{\nu_K}'}{q_{\mu_1} \dots q_{\mu_K}}\tr[\prs{\xi_m} M_{\alpha_K}\dots M_{\alpha_1} \prs{\psi_l} M_{\alpha_1}^\dagger \dots M_{\alpha_K}^\dagger] \,  p_m'.
\end{align}
Consequently, by combining \eq{eq:ent-prod}, \eq{eq:forward-traj-prob-kraus}, and \eq{eq:reverse-traj-prob-kraus}, we obtain the following expression for the stochastic entropy production along each trajectory:
\begin{align}
\Delta_i s[\Gamma] &= \ln{\frac{P(\Gamma)}{\tilde P(\tilde \Gamma)}}, \nonumber \\
&=  \ln{\frac{p_l}{p_m'}} + \sum_{k=1}^K \ln{\frac{q_{\mu_k}}{q_{\nu_k}'}}.
\end{align}
In other words, the stochastic entropy production is purely a function of the  probability distributions for the system and bath, at the beginning and end of the joint dynamics during the time intervals $(t_{k-1}, t_k)$.

Finally, we note that the entropy production, averaged with respect to the probability distribution for the forward trajectories, gives
\begin{align}\label{eq:DiS2.1}
\<\Delta_is[\Gamma]\> &:= \sum_\Gamma P(\Gamma)\Delta_is[\Gamma], \nonumber \\
&= \Delta S\sub{\s} + \sum_{k=1}^K\Delta S\sub{\e}^k \geqslant  0,
\end{align}
where $\Delta S\sub{\s}:= S(\rho\sub{\s}(t_K)) - S(\rho\sub{\s})$, $\Delta S\sub{\e}^k:= S(\rho\sub{\e}(t_k)) - S(\rho\sub{\e})$, and $S(\rho):= -\tr[\rho \, \ln{\rho}]$ is the von-Neumann entropy of state $\rho$. The inequality in the final line, which can be interpreted as an expression of the second law of thermodynamics, is due to: (i) lack of initial correlations between system and environment at the start of each joint evolution; (ii) sub-additivity of the von-Neumann entropy \cite{Petz-QI}; and (iii) the fact that unitary evolution is unital, i.e., does not decrease the von-Neumann entropy \cite{Uhlmann-Stochasticity,Nakahara-Decoherence}. In other words, the second law is always satisfied for open system dynamics. A more detailed discussion about the second law for quantum systems can be found in \cite{CH27}.

The entropy production $\Delta_\text{i}s[\Gamma]$ also fulfills a fluctuation Theorem:
 \bb
 \moy{e^{-\Delta_\text{i} s[\Gamma]}} &:=& \sum_{\Gamma : P(\Gamma) > 0} P(\Gamma) \dfrac{{\tilde P}(\tilde\Gamma)}{P(\Gamma)} \nonumber\\
&=& 1 - \lambda. \label{eq:IFT}
 \ee
\noindent Here the sum is restricted to the trajectories $\Gamma$ for which $P(\Gamma)\neq 0$, and $\lambda \in [0,1)$ corresponds to the cumulated probabilities of the reverse trajectories $\tilde \Gamma$ for which the probability of the corresponding forward trajectory, $P(\Gamma)$, is zero. A situation in which $\lambda >0$ is referred to as ``absolute irreversibility'' as it is associated with a strictly positive entropy production \cite{MurashitaThesis,Funo15}. This typically arises for a system initially restricted to a subpart of its Hilbert space and then allowed to relax in the whole space. $\lambda$ vanishes when the initial state of the system, $\rho\sub{\s}$, has full rank, i.e., when $\rho\sub{\s}$ has $d\sub{\s}$ positive eigenvalues. An example of a full-rank state is a thermal state at finite temperature.  Eq.\eqref{eq:IFT} is known as the "Integral Fluctuation Theorem" \cite{Seifert05} (IFT) as it generates other famous fluctuation theorems when applied to specific situations. For instance, when the system and the reservoir are initialized in thermal states at the beginning of the forward and backward processes, one obtains the quantum version of the celebrated Jarzynski equality \cite{Tasaki00,Kurchan01}.

We stress here that the time reversed transformation is a virtual reference with which the forward process is compared, rather than an actual physical process that needs to be implemented. Indeed, the fluctuation theorem and and second law can be tested without having to implement such reverse processes as the averages appearing in Eqs.\eqref{eq:DiS2.1} and \eqref{eq:IFT} are over the forward trajectories.

Finally, we emphasize that the discussion can be extended to more general kind of measurements performed on the system and the environment (see Appendix B).

\section{First law of thermodynamics}
\label{s:FirstLaw}

\subsection{Internal energy}
In this section, we focus on the energy exchanges occurring during the quantum trajectories defined above. In this chapter, we define the internal energy of system $\s$ when it is in the pure state $\ket{\psi\sub{\s}}$ following \cite{Alexia-measurement-thermodynamics} by:
\bb 
U(\lambda) := \bra{\psi\sub{\s}}H\sub{\s}(\lambda)\ket{\psi\sub{\s}} \label{eq:Ulambda}
\ee

 We stress that here $\ket{\psi\sub{\s}}$ can be any state of $\hs$ and in particular does not have to be an energy eigenstate of $H\sub{\s}(\lambda)$. $U(\lambda)$ can be interpreted as the average of results of projective energy measurement performed on many copies of the system. However, if the system is in an energy eigenstate, $U(\lambda)$ corresponds to an eigenvalue of $H\sub{\s}(\lambda)$. As a consequence, if one includes projective measurements of the system's energy in the thermodynamic transformation under study, one finds that $U(\lambda)$ is given by the outputs of such measurements. This is the Two-Point Measurement (TPM) approach of quantum thermodynamics which led to pioneering results such as the first quantum fluctuation theorems \cite{Kurchan01,Mukamel03,Talkner07,Campisi11}(see also \cite{CH9}). However, definition \eqref{eq:Ulambda} can be seen as an extension of this conception of internal energy as it allows us to follow the variation in internal energy of the system even when no projective measurement is performed, provided the system's state is known. This definition is extremely useful in the context of quantum trajectories experiments which precisely lead to the knowledge of the system's quantum state at any time. Note that this definition \eqref{eq:Ulambda} can be extended in the case of a mixed state $\rho\sub{\s}$ to $U(\lambda) := \tr[\rho\sub{\s}H\sub{\s}]$. This definition of the internal energy was used in seminal studies of open quantum system thermodynamics \cite{Alicki79,intro}.

\subsection{Work and heat: two historic approaches}

Just like in classical thermodynamics, we are now interested in splitting the variations of $U(\lambda)$ into work, a deterministic form of energy that can be readily used for many purposes in a controllable way, and heat -- a stochastic/uncontrollable form of energy exchange. The definitions of heat and work for quantum systems has a long history of attempts, mainly based on two strategies for which we briefly sum up the arguments in the following two subsections, in order to introduce the unifying paradigm we will use in this chapter.

\subsubsection{"Heat first"}

Studies focusing on the cases where the environment is a thermal reservoir have generally defined heat from the outcomes of energy measurements performed on the thermal reservoir. In the context of quantum trajectories, this corresponds to an unraveling in which the bases $\{\ket{\varphi_n}\}$ and $\{\ket{\phi_m}\}$ are both the reservoir's energy eigenbasis.  

The QJ unraveling presented above is generally the best suited to monitor this heat exchange, that we denote $Q_{\text{cl}}$ in the following: indeed, the outcomes are directly linked to a change of the number of excitations in the environment, and therefore to a variation of the reservoir's energy \cite{Hekking13}. Obtaining outcome $\alpha_0$ at time $t_k$ corresponds to a zero heat, $\delta Q_\text{cl}(t_k,\alpha_0)=0$, while each outcome $\alpha_j$ with $j\geqslant 1$ is associated with an amount of heat $\delta Q_\text{cl}(t_k,\alpha_j)$ given by the quantum of energy carried by the exchanged excitation. In the example of photon-counting, this can be summed up by saying that the heat exchanged with the reservoir is the energy carried by the counted photons. In the case of a reservoir at thermal equilibrium, the probabilities of picking each energy eigenstates in the forward and reverse processes, $q_\nu$ and $q_\mu'$, are identified with Boltzmann's distributions, such that $\moy{\delta Q_{\text{cl}}(t_k,\alpha_k)}$ is linked to the variation of the thermal reservoir's Von Neumann entropy according to $\Delta S_\e = - \moy{\delta Q_{\text{cl}}(t_k,\alpha_k)}/T$ \cite{CH9}.

This approach has been first used in the context of the TPM approach to quantum statistical mechanics, in which the change of the environment internal energy is assumed to be inferred from two projective measurements performed at the  beginning and at the end of the trajectory \cite{Campisi11}. Another formulation can be found in the so-called ``full counting statistics'' method \cite{Esposito09}, which consists of recording in a generating function the outputs of (fictitious) projective energy measurements performed on the environment at every time step. The heat statistics given by such a method is actually the same as obtained from the QJ unraveling \cite{Elouard18Fluo}. In such studies, the rest of the internal energy variations $\Delta U - Q_\text{cl}$ was identified as the work.

\subsubsection{"Work first"}

Another approach, in its spirit closer to the initial formulation of thermodynamics, consists in identifying the energy deterministically exchanged with the system as the work. In the case where the dynamics is described by a Lindblad master equation, the work should then appear in the unitary part of the dynamics, while heat, which is uncontrollable, should come from the non-unitary terms. It turns out that non-zero contributions of the unitary terms in Eq.\eqref{eq:Lindblad} to the  variations of $U(\lambda)$ solely appears when the parameter $\lambda$ is varied along the transformation\footnote{Note that we consider here the work performed on the total system $\s$. If $\s$ is composed of several subsystems, one can find deterministic energy exchanges within this subsystems, even when the total Hamiltonian of $\s$ is time-independent}. This variation can be interpreted as originating from an external driving field whose dynamics generates a sequence $\{\lambda_t\}_{t_i\leqslant t \leqslant t_f}$. One can then identify the elementary amount of work provided by this driving field to the system at time $t_k$ with (see \cite{Alicki79,intro})
\bb\label{eq:dW}
\delta W(\alpha_k,\lambda_{t_k}) = dt \dfrac{d\lambda_{t}(t_k)}{dt} \bra{\Psi_\Gamma(t_k)}\frac{d}{d\lambda}H\sub{\s}(\lambda_{t_k})\ket{\Psi_\Gamma(t_k)} .
\ee

Note that the increment of work does not depend on  the measurement outcome $\alpha_k$ obtained at time $t_k$, which confirms the deterministic nature of the energy exchange. Then, the remaining terms in the variation of $U(\lambda)$ are classified as heat, acknowledging their uncontrollable nature.    
This approach has the great advantage to be compatible with any Lindblad equation (even for non-thermal dissipation) and any unraveling.

\subsection{The notion of quantum heat}\label{sec:Qq}

Recent studies exploring the impact of definition \eqref{eq:Ulambda} to states which are not energy eigenstates \cite{Alexia-measurement-thermodynamics} have evidenced that these two approaches do not agree in general. Focusing on the case of the environment being a thermal reservoir, it turns out that the heat computed as the change of the reservoir's energy and the work provided by the external drive computed from Eq.\eqref{eq:dW} do not in general add up to the total variations of $U(\lambda)$. This mismatch  solely appears when coherences are built by the external drive, or are initially present in the system's state, in the basis in which the reservoir induces jumps (in the case of a QJ unraveling) or more generally in the basis in which the measurement scheme induces decoherence.

This mismatch does not mean that energy conservation fails, but simply that there is a flow of energy that is uncontrolled (non-unitary) while not provided by the thermal reservoir. This flow of energy can either be provided by the driving field (and in contrast to work $W$ occur in a stochastic way because it is triggered by the interaction with the reservoir) or by other sources not explicitly included in the quantum description, such as the measuring apparatus involved in the unraveling of the Lindblad equation. We stress that the precise origin of this energy flow can always be determined for a given problem from a careful microscopic model of the driving fields and measuring apparatuses involved in all steps of the studied protocol (including preparation steps) \cite{ElouardThesis}.

A dramatic example is the case in which the Lindblad equation is generated by the continuous monitoring of some observable (the environment is then a meter coupled to the system and periodically measured, see Section \ref{s:Ex3}). In this situation, there is no thermal reservoir, but still some non-unitary energy flow exchanged with the system, which is not of thermal origin (there is no temperature involved) but which should not be called work because of its uncontrolled nature. In Refs. \cite{Elouard2017a,Yi17, Elouard18}, this measurement-induced heat flow is exploited to build engines solely fueled by the observation process, a remarkable specificity of the quantum world.

In the end of this chapter, we shall follow \cite{Alexia-fluctuation-engineered-reservoir,Elouard18,Elouard18Fluo,Alexia-measurement-thermodynamics,Gherardini18} and call this quantity of purely quantum origin ``quantum heat''. We will denote $\delta Q_\text{q}(t_k,\alpha_k)$ the quantum heat increment at time $t_k$.

The expression of the quantum heat (just as the expression of the classical, thermal heat $Q_\text{cl}$) depends on the unraveling, and it may not always be possible to split the total incoherent energy exchange into thermal and quantum heat. We analyze in Sections \ref{s:Ex1}-\ref{s:Ex3} examples in which this splitting is possible 

Finally the first law at the single trajectory level reads as
\bb
\Delta U(\Gamma) = W(\Gamma) + Q_\text{cl}(\Gamma) + Q_\text{q}(\Gamma),
\ee
where the trajectory dependent quantities $X(\Gamma)$, for $X = W,\; Q_\text{cl},\; Q_\text{q}$ are computed by summing all the elementary increments $\delta X(t_k,\alpha_k)$ along trajectory $\Gamma$.

\section{Example 1: Quantum and classical entropy production due to thermalization}\label{s:Ex1}
Here we shall study the simplest example of interest where a system is allowed to relax to thermal equilibrium by interacting with a thermal bath at temperature $T$. When the system possesses initial coherences with respect to its energy eigenbasis, the resulting entropy production can be decomposed into a quantum and classical component, with concomitant fluctuations in classical and quantum heat, respectively. This will largely follow the work presented in \cite{Mohammady18}. 

Let the system $\s$, with the Hamiltonian $H\sub{\s} = \sum_i e_i \prs{e_i}$, be prepared in the state $\rho\sub{\s} = \sum_l p_l \prs{\psi_l}$. The bath, on the other hand, has the Hamiltonian $H\sub{\e} = \sum_i \epsilon_i \pre{i}$, and is initially in the canonical state $\tau\sub{\e}:= e^{-H\sub{\e}/k_B T}/ Z\sub{\e}$, where $Z\sub{\e}:= \tr[e^{-H\sub{\e}/k_B T}]$ is its partition function.  The thermalization process is effected by a joint unitary evolution $V$, which commutes with the total Hamiltonian $H\sub{\s} + H\sub{\e}$, such that $\tr\sub{\e}[V(\rho\sub{\s}\otimes \tau\sub{\e})V^\dagger] = \tau\sub{\s}:= e^{-H\sub{\s}/k_B T}/ Z\sub{\s}$ and $\tr\sub{\s}[V(\rho\sub{\s}\otimes \tau\sub{\e})V^\dagger] = \tau\sub{\e}'$. We shall assume that the bath is sufficiently large so that $\tau\sub{\e}'$ will be ``almost'' thermal. This is quantified by the relative entropy $D[\tau\sub{\e}'\| \tau\sub{\e}]:= \tr[\tau\sub{\e}' (\ln{\tau\sub{\e}'} - \ln{\tau\sub{\e}})]$ being vanishingly small. 

As in the general example discussed in \sect{s:Qtraj-open-system}, we may construct the quantum trajectories by measuring the system and environment, before and after the joint unitary evolution, with observables that do not disturb the subsystems. However, we may ``augment'' these trajectories by also performing energy measurements on the system prior to its interaction with the thermal reservoir. This is valid as the resulting average entropy production, and average internal energy change of the system, will stay the same. The augmented trajectories correspond to the TPM protocol introduced above which has been historically used to derive the first versions of the quantum fluctuation theorems \cite{Kurchan01,Mukamel03,Talkner07}. The present analysis allows us to quantify the thermodynamic consequences of the initial energy measurement. 

The trajectories of the system, containing the energy measurement prior to thermalization, are defined as $\Gamma_\s:= (l,m,n)\equiv \ket{\psi_l} \mapsto \ket{e_m} \mapsto \ket{e_n}$. The trajectories of the bath, meanwhile, are $\Gamma_\e := (\mu, \nu) \equiv \ket{\mu} \mapsto \ket{\nu}$. The joint trajectory $\Gamma := ((l,m,n), (\mu, \nu))$ can therefore be decomposed into a ``decoherence'' trajectory $\Gamma_\mathrm{q}:= (l,m) \equiv \ket{\psi_l} \mapsto \ket{e_m}$, which only pertains to the system,  followed by a ``classical thermalization'' trajectory $\Gamma_\mathrm{cl}:=((m,n),(\mu,\nu))$ which is the joint trajectory of system and bath during thermalization. 
As the system Hamiltonian is time-independent, the variation of its internal energy is only heat. The change in internal energy of the system during the decoherence trajectories is quantum heat as defined in Section \ref{sec:Qq}: 
\begin{align}
Q_\mathrm{q}(\Gamma) := \tr[H\sub{\s}(\prs{e_m} - \prs{\psi_l})].
\end{align}
 The change in internal energy of the system during the classical thermalization trajectories is classical heat which, due to the energy conservation of the joint unitary evolution, together with the fact that both system and bath start and end in energy eigenstates, equals the heat emitted by the bath:
\begin{align}
Q_\mathrm{cl}(\Gamma) := \tr[H\sub{\s}(\prs{e_n} - \prs{e_m})] = \tr[H\sub{\e}(\pre{\mu} - \pre{\nu})].
\end{align}
It is evident that the total change in the system's internal energy along a trajectory is $Q(\Gamma) = Q_\mathrm{q}(\Gamma) + Q_\mathrm{cl}(\Gamma)$.

Using the methods delineated in \sect{s:Qtraj-open-system} and \sect{s:reversed-traj-entropy-prod}, we may evaluate the probabilities for each sub-trajectory, as well as their time-reversed counterparts, to be
\begin{align}
P(\Gamma_\mathrm{q})&= \tr[\prs{e_m}\prs{\psi_l}] \, \tr[\prs{\psi_l}\rho\sub{\s}], 
&\tilde P(\tilde \Gamma_\mathrm{q})&=   \tr[\prs{e_m}\prs{\psi_l}] \, \tr[\prs{e_m}\eta\sub{\s}], \nonumber \\
P(\Gamma_\mathrm{cl})&= \tr[\prs{e_n}\Phi_{\mu,\nu}(\prs{e_m})] \,\tr[\prs{e_m}\eta\sub{\s}],
&\tilde P(\tilde\Gamma_\mathrm{cl}) & =  \tr[\prs{e_m}\tilde \Phi_{\mu,\nu}(\pr{e_n})] \, \tr[\prs{e_n}\tau\sub{\s}],
\end{align}
where 
\begin{align}
\eta\sub{\s}:= \sum_m \prs{e_m}\rho\sub{\s} \prs{e_m}
\end{align}
is the same as the initial state $\rho\sub{\s}$, except that its coherences with respect to the Hamiltonian eigenbasis have been removed.

The average quantum entropy production, therefore, is 
\begin{align}\label{eq:average quantum entropy production thermalization}
\<\Delta_is[\Gamma_\mathrm{q}]\> &:= \sum_{\Gamma_\mathrm{q}} P(\Gamma_\mathrm{q}) \ln{\frac{P(\Gamma_\mathrm{q})}{\tilde P(\tilde \Gamma_\mathrm{q})}}, \nonumber \\
& = \sum_{l,m} \tr[\prs{e_m}\prs{\psi_l}] \, \tr[\prs{\psi_l}\rho\sub{\s}] \ln{\frac{\tr[\prs{\psi_l}\rho\sub{\s}]}{\tr[\prs{e_m}\eta\sub{\s}]}}, \nonumber \\
&= S(\eta\sub{\s}) - S(\rho\sub{\s}) \equiv D[\rho\sub{\s} \| \eta\sub{\s}].
\end{align}
Moreover, noting that $\tilde \Phi_{\mu, \nu}(\rho) = (q_\nu'/q_\mu) \Phi^\dagger (\rho)$, where $q_\mu := \<\mu|\tau\sub{\e} |\mu\>$ and $q_\nu' := \<\nu| \tau\sub{\e}'|\nu\>$, while $\sum_\nu \tr[\Phi_{\mu,\nu}(\rho)] = q_\mu$ and $\sum_\mu \tr[\Phi_{\mu,\nu}(\rho)] = q_\nu'$, it follows that the average classical entropy production is given as 
\begin{align}
\<\Delta_is[\Gamma_\mathrm{cl}]\> &:= \sum_{\Gamma_\mathrm{cl}} P(\Gamma_\mathrm{cl}) \ln{\frac{P(\Gamma_\mathrm{cl})}{\tilde P(\tilde \Gamma_\mathrm{cl})}}, \nonumber \\
&= \sum_{m,n,\mu,\nu} \tr[\prs{e_n}\Phi_{\mu,\nu}(\prs{e_m})] \,\tr[\prs{e_m}\eta\sub{\s}] \left(\ln{\frac{q_\mu}{q_\nu'}} + \ln{\frac{\tr[\prs{e_m}\eta\sub{\s}]}{\tr[\prs{e_n}\tau\sub{\s}]}} \right), \nonumber \\
& = S(\tau\sub{\s}) - S(\eta\sub{\s}) + S(\tau\sub{\e}') - S(\tau\sub{\e}).
\end{align}

Note that, given a thermal state $\tau:= e^{-H/k_B T}/Z$, the following relation holds for all states $\rho$:
\begin{align}\label{eq:thermal-state-relent}
S(\rho) - S(\tau)  = \frac{1}{k_B T} \tr[H(\rho - \tau)] - D[\rho \| \tau].
\end{align}
Using \eq{eq:thermal-state-relent}, together with the fact that $\tr[H\sub{\s} (\rho\sub{\s} - \eta\sub{\s})]=0$, and that the thermalization unitary interaction $V$ conserves energy, it follows that the average classical entropy production is
\begin{align}\label{eq:average classical entropy production thermalization}
\<\Delta_is[\Gamma_\mathrm{cl}]\> & = S(\tau\sub{\s}) - S(\eta\sub{\s}) +S(\tau\sub{\e}')  - S(\tau\sub{\e}) , \nonumber \\
& = \frac{1}{k_B T} \left(\tr[H\sub{\s}(\tau\sub{\s} - \eta\sub{\s})] + \tr[H\sub{\e} (\tau\sub{\e}' - \tau\sub{\e})] \right) + D[\eta\sub{\s}\|\tau\sub{\s}] - D[\tau\sub{\e}' \| \tau\sub{\e}], \nonumber \\ 
&= D[\eta\sub{\s}\|\tau\sub{\s}] - D[\tau\sub{\e}' \| \tau\sub{\e}],  \nonumber \\
& \approx D[\eta\sub{\s} \| \tau\sub{\s}].
\end{align}
Finally, we note that the entropy production is additive along sub-trajectories, so that the average entropy production for the full process is $\<\Delta_is[\Gamma]\> =\<\Delta_is[\Gamma_\mathrm{q}]\> + \<\Delta_is[\Gamma_\mathrm{cl}]\> = D[\rho\sub{\s} \| \tau\sub{\s}]$.

As the relative entropy is a non-negative quantity that vanishes if and only if both arguments are identical, it follows that quantum entropy production vanishes only when the initial state has no coherences with respect to the Hamiltonian eigenbasis, i.e., when $\rho\sub{\s}=\eta\sub{\s}$. Note that while the average quantum heat $\<Q_\mathrm{q}(\Gamma)\>$ is always zero, the existence of its ``fluctuations'' witnesses the presence of quantum entropy production.

\section{Example 2: Photo-counting of fluoresence}
\label{s:Ex2}

We now analyze the example of the canonical situation of quantum optics: a two-level atom (qubit) coherently driven and coupled to its electromagnetic environment (a thermal reservoir at temperature $T$). We consider a photon-counting measurement, i.e. that the photons emitted and absorbed by the qubit are counted at intervals of $\Delta t$, with $\gamma \Delta t \ll 1$ where $\gamma$ is the spontaneous emission rate of the qubit. While implementing this measurement scheme at zero temperature only requires a photon-counter (absorptions impossible), its extension at non-zero temperature requires some tricks like reservoir engineering \cite{Horowitz12,Alexia-fluctuation-engineered-reservoir} or a finite-size reservoir \cite{Pekola13}. The present analysis follows from \cite{Alexia-fluctuation-engineered-reservoir,Elouard18Fluo}. Other analyses of the fluorescence thermodynamics focusing on the Floquet master equation, i.e. describing the dynamics of the atom coarse-grained over a time-scale much larger than a Rabi oscillation, can be found in \cite{Alicki13,Bulnes15,Donvil18}. Note that a similar situation has been experimentally implemented and analyzed with a QSD unraveling \cite{Naghiloo17}.

\begin{figure}[!htb]
\begin{center}
\includegraphics[width = 0.6\textwidth]{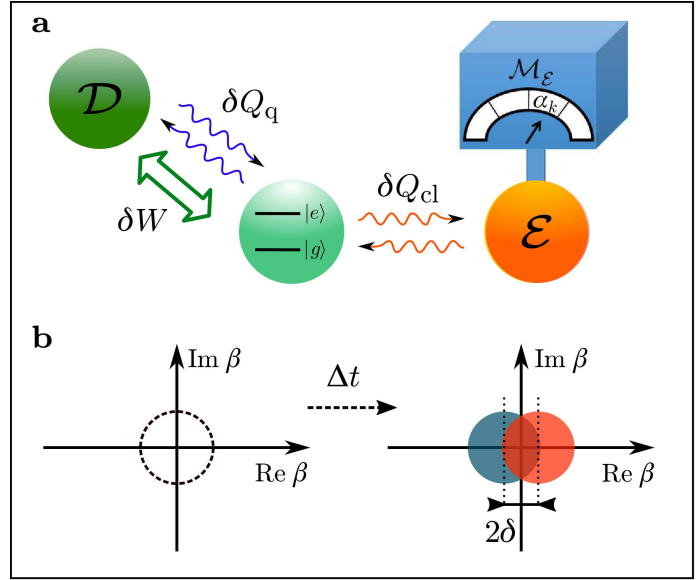}
\end{center}
\caption{\textbf{a}: Situation studied in Example 2. A qubit of state $\{\ket{e},\ket{g}\}$ is driven by an external drive ${\cal D}$ modeled via the time-dependent Hamiltonian $H_\text{d}(t)$, and coupled to a thermal bath $\cal E$. A measuring apparatus monitors the variations of the number of photons in $\cal E$ such that the systems follows trajectories corresponding to the quantum jump unraveling. Between $t$ and $t+\Delta t$, the drive exchanges energy with the qubit in two forms: the work $\delta W$ (coherent energy exchange) and the quantum heat $\delta Q_\text{q}$ (incoherent/stochastic). Besides, the thermal bath exchanges classical heat $\delta Q_\text{cl}$ with the qubit. \textbf{b}: Measurement scheme of Example 3: The cavity mode used as a meter is initially in the vacuum. It is then coupled to the qubit during $\Delta t$, which induces a positive (resp. negative) coherent displacement when the qubit is in the excited (resp. ground) state.}
\label{fig:2}
\end{figure}

The driven qubit is described by Hamiltonian $H\sub{\s}(t) = H_0 + H_\text{d}(t) = \hbar\omega_0/2\sigma_z + \hbar g (\sigma_-e^{i\omega_L t}+\sigma_+e^{-i\omega_L t})$, with $\omega_0$ the bare energy of the qubit, $\omega_L$ the frequency of the drive and $g$ the Rabi frequency (fixed by the drive intensity). We have introduced the Pauli matrix $\sigma_z = \ket{e}\bra{e}-\ket{g}\bra{g}$ and the qubit ladder operators $\sigma_- = \ket{g}\bra{e} = \sigma_+^\dagger$. The thermal reservoir is modeled as a collection of harmonic oscillators of Hamiltonian $H_\e = \sum_l \hbar\omega_l a_l^\dagger a_l$, $a_l$ being the annihilation operator in mode $l$. The coupling between the qubit and the reservoir is captured by Hamiltonian $H_\text{int} = \sum_l g_l(\sigma_- a_l^\dagger + \sigma_+ a_l)$, where the coupling strengths $g_l$ are taken as real without loss of generality. 

As soon as $g\ll \omega_0$ and the reservoir has a correlation-time $\tau_\text{c}$ much shorter than the inverses of $\gamma\equiv \sum_l g_l^2 \delta(\omega_l-\omega_0)$ and $g$, one can coarse-grain the unitary qubit-reservoir dynamics on a time-scale $\Delta t$ fulfilling $\tau_\text{c} \ll \Delta t \ll \gamma^{-1},g^{-1}$, and trace over the reservoir degrees of freedom to obtain a Lindblad equation for the qubit density operator \cite{cohen}, which has the form \eqref{eq:Lindblad}-\eqref{eq:Dj} with the Lindblad operators $\{L_j\}_{j=\pm}$:
\bb
L_- &=& \sqrt{\gamma(\bar n +1)}\sigma_-\\
L_+ &=& \sqrt{\gamma\bar n}\sigma_+,
\ee
\noindent where $\bar n = (e^{\hbar\omega_0/k_\text{B}T}-1)^{-1}$.
Alternatively, when the variation of the number of photons in the bath is monitored every $\Delta t$, the dynamics of the qubit is captured by a QJ unraveling as described in Section \ref{s:QJ}, with three Kraus operators labeled by $\alpha\in\{+,-,0\}$. $M_\pm = \sqrt{\Delta t}L_\pm$ are associated with the absorption and the emission of a photon, while $M_0 = 1-iH\Delta t - \tfrac{\Delta t}{2}(L_+^\dagger L_+ + L_-^\dagger L_-)$ is associated with the detection of no variation of the photon number in the environment between $t_k$ and $t_{k+1}$. We denote $\ket{\Psi_\Gamma(t_k)}$ the state of the qubit at time $t_k$ along trajectory $\Gamma = ((l,m),\vec\alpha)$.

\subsubsection*{First law}
We first analyze the energy exchanges along such QJ trajectories applying the results of Section \ref{s:FirstLaw}. The time-dependence of $H\sub{\s}$ due to the interaction with the classical external drive can be understood as a thermodynamical transformation in which the parameter $\lambda_t = \omega_L t$ is varied. For the sake of simplicity, we consider the TPM protocol which is associated with quantum Jarzynski equality \cite{Mukamel03,Campisi09}: The qubit is initially at thermal equilibrium  while the drive is off ($g=0$), and measured at time $t_i$ in its energy eigenbasis, yielding state $\ket{\psi_l}\in\{\ket{e},\ket{g}\}$ with respective probability $p_e^\text{(th)} = e^{-\hbar\omega_0/k_\text{B}T}/(1+e^{-\hbar\omega_0/k_\text{B}T})$ and $p_g^\text{(th)} = 1/(1+e^{-\hbar\omega_0/k_\text{B}T})$. Then, the drive is switched on up to time $t_f$, where another energy measurement is performed on the qubit yielding $\ket{\psi_m}\in\{\ket{e},\ket{g}\}$.

The elementary work performed on the qubit via the drive between times $t_{k-1}$ and $t_{k}$ can be deduced from Eq.\eqref{eq:dW}:
\bb
\delta W(\alpha_k,t_k) &=& \Delta t\bra{\Psi_\Gamma(t_k)}\partial_t H_\text{d}(t)\ket{\Psi_\Gamma(t_k)}\nonumber\\
&=& -g\Delta t \, \text{Im}(s_\Gamma(t_k)),
\ee
where $s_\Gamma(t_k) = \bra{\Psi_\Gamma(t_k)}\sigma_-\ket{\Psi_\Gamma(t_k)}e^{i\omega_Lt_k}$ is the atomic dipole in the frame rotating at the drive frequency, and $\ket{\Psi_\Gamma(t_k)}$ is the qubit's state at time $t_k$ along trajectory $\Gamma$. Meanwhile, one can identify the elementary heat exchanged with the thermal reservoir between time $t_{k-1}$ and $t_k$ as the energy carried by the exchanged photons:
\bb
\delta Q_\text{cl}(\alpha_k,t_k) = \alpha_k\hbar\omega_0, \quad\alpha_k\in\{+,-,0\}.
\ee
Finally, as the drive induces coherences in the bare atomic basis, that are erased when a photon is emitted or absorbed, there is a non-zero quantum heat contribution. This energy is provided by the same source as the work, i.e. the light driving the qubit, but in an uncontrolled/incoherent manner due to the interaction with the reservoir:
\bb
\delta Q_\text{q}(\alpha_k,t_k) = \left\{\begin{array}{ll}
-\hbar\omega_0 \dfrac{1+z_\Gamma(t_k)}2,& \alpha_k = +\nonumber\\
\hbar\omega_0  \dfrac{1-z_\Gamma(t_k)}2, &\alpha_k =-\nonumber\\
-\bar g \dfrac{\gamma (\bar n +1)}2 \Delta t \, \text{Re}(s_\Gamma(t_k)) - \hbar\omega_0\gamma\Delta t \vert s_{\Gamma}(t_k)\vert^2, &\alpha_k = 0.
\end{array}\right.
\ee
\noindent Here $z_\Gamma(t_k) =  \bra{\Psi_\Gamma(t_k)}\sigma_z\ket{\Psi_\Gamma(t_k)}$ is the qubit population at time $t_k$ along trajectory $\Gamma$.

One can evaluate the mean energy exchanges using the probability of each outcome $\alpha_k$ at time $t_k$: $p_{\alpha_k} = \bra{\Psi_\Gamma(t_k)}M_{\alpha_k}^\dagger M_{\alpha_k}\ket{\Psi_\Gamma(t_k)}$. It is interesting to look at the steady state value of the energy fluxes (denoted by the subscript $\infty$) obtained for $\gamma t \gg 1$, when the atom has relaxed in the steady state of the optical Bloch equations \cite{cohen}:
\bb
\moy{\dot W}_\infty &=&  \gamma\hbar\omega_L \dfrac{g^2}{2 g^2 + 4 \delta^2 + \gamma^2(2\bar n +1)^2}\\
\moy{\dot Q_\text{cl}}_\infty &=& -\gamma\hbar\omega_0  \dfrac{g^2}{2 g^2 + 4 \delta^2 + \gamma^2(2\bar n +1)^2}\\
\moy{\dot Q_\text{q}}_\infty &=&  \gamma\hbar\delta \dfrac{g^2}{2 g^2 + 4 \delta^2 + \gamma^2(2\bar n +1)^2},
\ee
with $\delta = \omega_0-\omega_L$ the detuning between the qubit and the driving frequencies. All three quantities vanish for $g=0$ or for a very large detuning $\vert\delta\vert \gg \gamma,g$ such that the qubit is effectively decoupled from the driving field. The average quantum heat vanishes at resonance ($\delta = 0$) while it still features non-zero fluctuations \cite{Alexia-fluctuation-engineered-reservoir}. The steady state first law reads $\moy{\dot W}_\infty+\moy{\dot Q_\text{cl}}_\infty+\moy{\dot Q_\text{q}}_\infty = 0$.

\subsubsection*{Second law and Jarzynski equality}

The time-reversal rules of Section \ref{s:reversed-traj-entropy-prod} lead, to the following set of time-reversed Kraus operators:
\bb
\tilde M_+ = M_- + {\cal O}(\gamma\Delta t)^2\nonumber\\
\tilde M_- = M_+ + {\cal O}(\gamma\Delta t)^2\nonumber\\
\tilde M_0 = M_0^\dagger + {\cal O}(\gamma\Delta t)^2.
\ee
As these Kraus operators fulfill the relation $\tilde M_{\alpha_k} = e^{\delta Q_\text{cl}(\alpha_k,t_k)/2k_\text{B}T}M_{\alpha_k}^\dagger$, the probabilities of the forward and time-reversed trajectories are simply related by the following relation, which is sometimes called the Detailed Fluctuation Theorem:
\bb
\tilde P(\tilde \Gamma) = \dfrac{p'_m}{p_l}e^{Q_\text{cl}(\Gamma)/k_\text{B}T}P(\Gamma).
\ee
This in turn entails the following form for the stochastic entropy production:
\bb
\Delta_\text{i}s[\Gamma] = \ln{p_l}-\ln{p'_m}- \dfrac{Q_\text{cl}(\Gamma)}{k_\text{B}T}.
\ee
This entropy production fulfill the IFT Eq.\eqref{eq:IFT} and the second law:
\bb
\moy{\Delta_\text{i}s[\Gamma]} = \Delta S\sub{\s}- \dfrac{\moy{Q_\text{cl}(\Gamma)}}{k_\text{B}T} \geqslant 0.
\ee
Under the assumption that both the initial and final distribution of qubit states $p_l$ and $p_m'$ are drawn from thermal equilibrium populations $\{p_e^\text{(th)},p_g^\text{(th)}\}$, the entropy production takes the form
\bb
\Delta_\text{i}s[\Gamma] &=& \dfrac{1}{k_\text{B}T}\left(\Delta U\sub{\s}(\Gamma) -Q_\text{cl}(\Gamma)- \Delta F\right) \nonumber\\
&=& \dfrac{1}{k_\text{B}T}\left(W(\Gamma)  +Q_\text{q}(\Gamma)- \Delta F\right),
\ee
such that the IFT becomes the quantum version of the well-known Jarzynski equality. In the present example $\Delta F = k_\text{B}T(\log Z(t_f)-\log Z(t_i)) = 0$ as the initial and final state are drawn from the same thermal distribution. Note the presence of the quantum heat term in the entropy production, which is a witness of a genuinely quantum contribution to the entropy production in this transformation: the continuous erasure by the qubit's environment of the coherences induced by the drive in the bare qubit basis. The results of Ref.\cite{Hekking13} are retrieved by identifying the total energy provided by the drive with the sum of the work $W$ and quantum heat $Q_\text{q}$.

\section{Example 3: continuous measurement of a qubit's observable}
\label{s:Ex3}

\subsubsection*{Kraus operators}
We consider a setup implementing the continuous measurement of the observable $\sigma_z$ of a qubit whose Hamiltonian is $H\sub{\s} = \hbar\omega_0\sigma_z$. This can be done by coupling the qubit to a cavity of Hamiltonian $H_\e = \hbar\omega_c b^\dagger b$, where $b$ is the annihilation operator in the cavity mode. It was recently demonstrated that a very efficient technique consists in exploiting the longitudinal coupling \cite{Didier15} which can be modeled by Hamiltonian $H_\text{int} = -i\sqrt{\gamma_\text{m}/\Delta t}(b^\dagger + b)\sigma_z$, where $\gamma_\text{m}$ is the measurement strength. For the sake of simplicity, we consider a transient measurement corresponding to a finite-time qubit-cavity interaction rather than the readout of the steady-state cavity proposed in \cite{Didier15} which implies a non-unitary evolution of the cavity. The cavity field is initially in the vacuum state $\ket{0}$. After interacting during $\Delta t$ with the qubit, the field is displaced on the real axis by an amplitude proportional to the expectation value of $\sigma_z$. This value can then be deduced from a homodyne measurement of the real quadrature of the field. In this example, we consider the weak measurement limit, i.e. $\gamma_\text{m}\Delta t\ll 1$ such that the two possible final coherent states of the cavity mode are strongly overlapping (see Fig.\ref{fig:2}\textbf{b}). Note that a very similar situation exploiting a dispersive coupling between the qubit and the cavity has recently been implemented to experimentally demonstrate fluctuation theorems in presence of quantum measurement and feedback \cite{Naghiloo18}.

The Kraus operator associated with finding the cavity field in state $\ket{\beta}$ after the interaction during $\Delta t$ is (in the interaction picture):
\bb
M(\beta) = \frac{1}{\sqrt\pi}\bra{\beta}e^{-i\Delta t H_\text{int} }\ket{0}.
\ee
Here we have taken the probability for the initial state of the cavity to be $q_{0} = 1$. We stress that despite the fact that $\{\ket{\beta}\}$ is an over-complete basis of $\he$, it still defines a proper Kraus decomposition of the quantum map under study, provided the factor $\dfrac{1}{\sqrt\pi}$ is included in the definition of $M(\beta)$, so as to ensure the normalization of the trajectory probabilities  \cite{Chantasri16}. This Kraus operator can be rewritten introducing the measurement outcome $I = -\text{Re}(\beta) /\sqrt{\gamma_\text{m}\Delta t}$, and $y = \text{Im}(\beta)$, and noting the interaction induces a displacement of the cavity's state by of amplitude $\sqrt{\gamma_\text{m}\Delta t}\sigma_z$:

\bb
M(y,I) = \dfrac{(\gamma_\text{m}\Delta t)^{1/4}}{\sqrt\pi}e^{-y^2/2-iy\sqrt{\gamma_\text{m}\Delta t}\sigma_z}e^{-\frac{\gamma_\text{m}\Delta t}{2}(I-\sigma_z)^2}e^{-i\omega_0\Delta t\sigma_z/2}
\ee

The probability law for outcome $(x,I)$ when the qubit is in state $\ket{\psi(t)} = c_e \ket{e}+ c_g \ket{g}$ reads
\bb\label{ex3:PyI}
P(y,I) = \dfrac{(\gamma_\text{m}\Delta t)^{1/2}}{\pi}e^{-y^2}\left(e^{-\gamma_\text{m}\Delta t(I-1)^2}\vert c_e\vert^2+e^{-\gamma_\text{m}\Delta t(I+1)^2}\vert c_g\vert^2\right),
\ee
and is composed of two Gaussian functions centered around $I = \pm 1$. In the weak measurement regime under study, the two Gaussian functions strongly overlap. We can check from \eqref{ex3:PyI} that $I$ follows a Gaussian law. It is therefore relevant to introduce a Wiener increment to describe the fluctuations of $I$ around its mean value $z = \bra{\psi}\sigma_z\ket{\psi}$:
\bb
I = z + \dfrac{dw(t)}{2\sqrt{\gamma_\text{m}}\Delta t}.
\ee
This allow to express $M(y,I)$ under a form similar to Eq.\eqref{eq:MQSD}, highlighting that the present measurement scheme corresponds to a QSD unraveling:
\bb
M(y,I) = \frac{e^{-y^2/2-iy\sqrt{\gamma_\text{m}\Delta t}\sigma_z}}{\pi^{1/4}} \left(\frac{e^{-dw(t)^2/2\Delta t}}{2 \pi \Delta t}\right)^{1/2}\left(1-\frac{\Delta t}{2}(\gamma_\text{m}+i\omega_0)+\sqrt{\gamma_\text{m}}dw(t)\sigma_z\right).
\ee
The corresponding Lindblad equation can be retrieved by averaging the conditioned density operator $M(y,I)\rho(t)M(y,I)^\dagger$ over the measurement outcomes:
\bb\label{eq:LindbladEx3}
\partial_t \rho = -i[H\sub{\s},\rho]+2\gamma_\text{m}(\sigma_z \rho\sigma_z - \rho).
\ee
This master equation captures the pure dephasing in the free energy eigenbasis of the qubit induced by the measurement. Note that the uncertainty on the imaginary part of the field (responsible for the dependence on $y$ of the probability distribution $P(y,I)$) cause the effective pure dephasing rate to be equal to twice the measurement rate. The pure dephasing can be decreased using a squeezed vacuum state \cite{Didier15}, but will remain larger than or equal to the measurement rate.

\subsubsection*{First law}
As is clear from Eq.\eqref{eq:LindbladEx3}, the qubit does not receive any work from an external agent, solely heat due to the non-unitary term proportional to $\gamma_\text{m}$. This heat is quantum heat as no thermal reservoir is involved in the problem. The quantum heat increment between $t_k$ and $t_{k+1}$ amounts to:
\bb\label{ex3:dQq}
\delta Q_\text{q}(y_k,I_k,t_k) &=& \dfrac{\moy{M(y_k,I_k)^\dagger H\sub{\s} M(y_k,I_k)}}{\moy{M(y_k,I_k)^\dagger M(y_k,I_k)}} - \moy{H\sub{\s}}\nonumber\\
&=& \hbar\omega_0\left(\sqrt{\gamma_\text{m}}dw(t)- 4\gamma \Delta t z_\Gamma(t)\right)(1-z_\Gamma(t)^2),
\ee
with $z_\Gamma(t) = \bra{\psi_\Gamma(t)}\sigma_z\ket{\psi_\Gamma(t)}$ the expectation value of $\sigma_z$ in the qubit's state at time $t$ along trajectory $\Gamma$. It is clear that the quantum heat is zero when the qubit is in state $\ket{e}$ ($z_\Gamma(t) = 1$) or $\ket{g}$ ($z_\Gamma(t) = -1$). As the qubit is always in a pure state along trajectory $\Gamma$, $1-z_\Gamma(t)^2 = \vert\bra{\psi_\Gamma(t)}e\rangle\langle g \ket{\psi_\Gamma(t)}\vert^2$ is directly related to the amplitude of the coherences between the energy eigenstates. Therefore, the quantum heat is non-zero solely if the qubit's state carries coherences in the $\{\ket{e},\ket{g}\}$ basis, i.e. solely when the measurement has an effect on the qubit's state. The quantum heat increment averaged over the measurement records (using probability $P(y_k,I_k)$ given by Eq.\eqref{ex3:PyI} of getting outcome $(y_k,I_k)$) is zero in this example. This result can also be deduced from the master equation noting that $\tr[H\sub{\s}\partial_t \rho\sub{\s}] = 0$ which implies that the average internal energy of the qubit is unchanged during the process. This is expected as the measured observable commutes with the Hamiltonian. Note however that, as it can be seen from Eq.\eqref{ex3:dQq}, the quantum heat takes non-zero values along single trajectories such that there are fluctuations around the average. Moreover, the average would be non-zero in the case of a measurement of an observable which does not commute with the Hamiltonian, allowing e.g. to fuel an engine \cite{Elouard2017a,Elouard18}.

\subsubsection*{Second law}

The time-reversed operator can be deduced from Eq.\eqref{eq:Mtilde}:
\bb
\tilde M(\beta) = \sqrt{P(\beta)}M(\beta)^\dagger,
\ee
using that the probability $P(\beta)$ for the cavity to start in state $\beta$ in the reversed trajectory, which is simply the probability of getting outcome $\beta$ in the direct trajectory. We deduce the entropy produced between $t_{k-1}$ and $t_{k}$:
\begin{align}
\delta_\text{i}s(y_k,I_k,t_k) &:= -\ln{P(y_k,I_k)} \nonumber \\ 
&= \ln{\frac{\pi}{\sqrt{\gamma_\text{m}\Delta t}}} + y^2 + \ln{e^{-\gamma\Delta t(I_k-1)^2}\vert c_e(t_k)\vert^2+e^{-\gamma(\Delta t +1)^2}\vert c_g(t_k)\vert^2}.
\end{align}

The entropy production contains a constant term scaling as $\ln{\sqrt{\gamma_\text{m}\Delta t}}$, which is in general very large, but finite\footnote{$\Delta t$ is fixed by the time-resolution of the cavity readout.}. This term can be interpreted by noting that the measurement generates a set of outcomes with typical resolution $\sqrt{\gamma_\text{m}\Delta t}$, such that one cavity state (the vacuum) is mapped onto a number of states scaling as $1/\sqrt{\gamma_\text{m}\Delta t}$ (see also section I-B of \cite{CH13}).

\section{Conclusion}
In this chapter, we have presented a formalism based on quantum trajectories to describe the stochastic thermodynamics of quantum open systems. We have introduced the notion of time-reversed quantum trajectories to compute the entropy production at the single trajectory level. This quantity fulfills the Integral Fluctuation Theorem and the second law. We have then analyzed the energy exchanges occurring during a single realization of a thermodynamic transformation to identify work and heat contributions to the variation of the internal energy of a quantum system. In particular, we have highlighted a quantum contribution to the heat received by the system which is related to the decoherence induced by the environment. 
Finally, we have illustrated this framework with three examples: the thermalization of a quantum system; the fluorescence of a driven qubit; and the continuous measurement of a qubit's observable. 

We have introduced the formalism of quantum stochastic thermodynamics focusing on an ideal situation involving a Markovian environment, monitored with perfect efficiency. However, the quantum trajectory framework is very flexible and can be extended to account for more complex situations. For instance, detection inefficiency can be introduced \cite{WisemanBook} to study its influence on quantum thermodynamics \cite{Wachtler16,Alexia-fluctuation-engineered-reservoir}. Besides, quantum trajectories can also encompass non-Markovianity \cite{Diosi97,Strunz99,Jack00,Jing13}, opening avenues towards characterization of non-Markovian quantum thermodynamics. 
Finally, feedback mechanisms exploiting the measurement record can be analyzed in the quantum trajectory framework, allowing e.g. to check generalizations of the quantum fluctuation theorems \cite{Naghiloo17,Naghiloo18}(see also \cite{CH9} Section III).

\section*{Ackowledgments}
C.E. acknowledges support by the US Department of Energy grant No. DE-SC0017890 and thanks Chapman University and the Institute for Quantum Studies for hospitality during this project.

\bibliography{References}

\begin{thebibliography}{80}
\providecommand{\natexlab}[1]{#1}
\providecommand{\url}[1]{\texttt{#1}}
\expandafter\ifx\csname urlstyle\endcsname\relax
  \providecommand{\doi}[1]{doi: #1}\else
  \providecommand{\doi}{doi: \begingroup \urlstyle{rm}\Url}\fi

\bibitem[M{\o}lmer et~al.(1993)M{\o}lmer, Castin, and Dalibard]{Molmer93}
Klaus M{\o}lmer, Yvan Castin, and Jean Dalibard.
\newblock Monte carlo wave-function method in quantum optics.
\newblock \emph{Journal of the Optical Society of America B}, 10\penalty0
  (3):\penalty0 524--538, March 1993.
\newblock ISSN 0740-3224.
\newblock \doi{10.1364/josab.10.000524}.
\newblock URL \url{http://dx.doi.org/10.1364/josab.10.000524}.

\bibitem[Wiseman(1996)]{Wiseman96}
H.~M. Wiseman.
\newblock Quantum trajectories and quantum measurement theory.
\newblock \emph{Quantum and Semiclassical Optics: Journal of the European
  Optical Society Part B}, 8\penalty0 (1):\penalty0 205--222, February 1996.
\newblock ISSN 1355-5111.
\newblock \doi{10.1088/1355-5111/8/1/015}.
\newblock URL \url{http://dx.doi.org/10.1088/1355-5111/8/1/015}.

\bibitem[Brun(2002)]{Brun02}
Todd~A. Brun.
\newblock A simple model of quantum trajectories.
\newblock \emph{American Journal of Physics}, 70\penalty0 (7):\penalty0
  719--737, July 2002.
\newblock ISSN 0002-9505.
\newblock \doi{10.1119/1.1475328}.
\newblock URL \url{http://dx.doi.org/10.1119/1.1475328}.

\bibitem[Nagourney et~al.(1986)Nagourney, Sandberg, and Dehmelt]{Nagourney86}
Warren Nagourney, Jon Sandberg, and Hans Dehmelt.
\newblock Shelved optical electron amplifier: Observation of quantum jumps.
\newblock \emph{Physical Review Letters}, 56\penalty0 (26):\penalty0
  2797--2799, June 1986.
\newblock ISSN 0031-9007.
\newblock \doi{10.1103/physrevlett.56.2797}.
\newblock URL \url{http://dx.doi.org/10.1103/physrevlett.56.2797}.

\bibitem[Sauter et~al.(1986)Sauter, Neuhauser, Blatt, and Toschek]{Sauter86}
Th~Sauter, W.~Neuhauser, R.~Blatt, and P.~E. Toschek.
\newblock Observation of quantum jumps.
\newblock \emph{Physical Review Letters}, 57\penalty0 (14):\penalty0
  1696--1698, October 1986.
\newblock ISSN 0031-9007.
\newblock \doi{10.1103/physrevlett.57.1696}.
\newblock URL \url{http://dx.doi.org/10.1103/physrevlett.57.1696}.

\bibitem[Hulet et~al.(1988)Hulet, Wineland, Bergquist, and Itano]{Hulet86}
Randall~G. Hulet, D.~J. Wineland, J.~C. Bergquist, and Wayne~M. Itano.
\newblock Precise test of quantum jump theory.
\newblock \emph{Physical Review A}, 37\penalty0 (11):\penalty0 4544--4547, June
  1988.
\newblock ISSN 0556-2791.
\newblock \doi{10.1103/physreva.37.4544}.
\newblock URL \url{http://dx.doi.org/10.1103/physreva.37.4544}.

\bibitem[Gleyzes et~al.(2007)Gleyzes, Kuhr, Guerlin, Bernu, Del\'{e}glise,
  Busk~Hoff, Brune, Raimond, and Haroche]{Gleyzes07}
S\'{e}bastien Gleyzes, Stefan Kuhr, Christine Guerlin, Julien Bernu, Samuel
  Del\'{e}glise, Ulrich Busk~Hoff, Michel Brune, Jean-Michel Raimond, and Serge
  Haroche.
\newblock Quantum jumps of light recording the birth and death of a photon in a
  cavity.
\newblock \emph{Nature}, 446\penalty0 (7133):\penalty0 297--300, March 2007.
\newblock ISSN 0028-0836.
\newblock \doi{10.1038/nature05589}.
\newblock URL \url{http://dx.doi.org/10.1038/nature05589}.

\bibitem[Murch et~al.(2013)Murch, Weber, Macklin, and Siddiqi]{Murch13}
K.~W. Murch, S.~J. Weber, C.~Macklin, and I.~Siddiqi.
\newblock Observing single quantum trajectories of a superconducting quantum
  bit.
\newblock \emph{Nature}, 502\penalty0 (7470):\penalty0 211--214, October 2013.
\newblock ISSN 0028-0836.
\newblock \doi{10.1038/nature12539}.
\newblock URL \url{http://dx.doi.org/10.1038/nature12539}.

\bibitem[Campagne-Ibarcq et~al.(2013)Campagne-Ibarcq, Flurin, Roch, Darson,
  Morfin, Mirrahimi, Devoret, Mallet, and Huard]{Campagne13}
P.~Campagne-Ibarcq, E.~Flurin, N.~Roch, D.~Darson, P.~Morfin, M.~Mirrahimi,
  M.~H. Devoret, F.~Mallet, and B.~Huard.
\newblock Persistent control of a superconducting qubit by stroboscopic
  measurement feedback.
\newblock \emph{Phys. Rev. X}, 3\penalty0 (2):\penalty0 021008+, May 2013.
\newblock ISSN 2160-3308.
\newblock \doi{10.1103/physrevx.3.021008}.
\newblock URL \url{http://dx.doi.org/10.1103/physrevx.3.021008}.

\bibitem[de~Lange et~al.(2014)de~Lange, Rist\`{e}, Tiggelman, Eichler,
  Tornberg, Johansson, Wallraff, Schouten, and DiCarlo]{deLange14}
G.~de~Lange, D.~Rist\`{e}, M.~.~J. Tiggelman, C.~Eichler, L.~Tornberg,
  G.~Johansson, A.~Wallraff, R.~.~N. Schouten, and L.~DiCarlo.
\newblock Reversing quantum trajectories with analog feedback.
\newblock \emph{Physical Review Letters}, 112\penalty0 (8), February 2014.
\newblock ISSN 0031-9007.
\newblock \doi{10.1103/physrevlett.112.080501}.
\newblock URL \url{http://dx.doi.org/10.1103/physrevlett.112.080501}.

\bibitem[Alicki(1979)]{Alicki79}
R.~Alicki.
\newblock The quantum open system as a model of the heat engine.
\newblock \emph{Journal of Physics A: Mathematical and General}, 12\penalty0
  (5):\penalty0 L103--L107, May 1979.
\newblock ISSN 0305-4470.
\newblock \doi{10.1088/0305-4470/12/5/007}.
\newblock URL \url{http://dx.doi.org/10.1088/0305-4470/12/5/007}.

\bibitem[Seifert(2008)]{Seifert08}
U.~Seifert.
\newblock Stochastic thermodynamics: principles and perspectives.
\newblock \emph{The European Physical Journal B}, 64\penalty0 (3-4):\penalty0
  423--431, 2008.
\newblock \doi{10.1140/epjb\%252fe2008-00001-9}.
\newblock URL \url{http://dx.doi.org/10.1140/epjb\%252fe2008-00001-9}.

\bibitem[Sekimoto(2010)]{Sekimoto10}
K.~Sekimoto.
\newblock \emph{Stochastic energetics}.
\newblock Springer, 2010.
\newblock ISBN 9783642054112.
\newblock URL \url{http://www.worldcat.org/isbn/9783642054112}.

\bibitem[Jarzynski(1997)]{Jarzynski97}
C.~Jarzynski.
\newblock Nonequilibrium equality for free energy differences.
\newblock \emph{Phys. Rev. Lett.}, 78\penalty0 (14):\penalty0 2690--2693, April
  1997.
\newblock ISSN 0031-9007.
\newblock \doi{10.1103/physrevlett.78.2690}.
\newblock URL \url{http://dx.doi.org/10.1103/physrevlett.78.2690}.

\bibitem[Crooks(1999)]{Crooks99}
Gavin~E. Crooks.
\newblock Entropy production fluctuation theorem and the nonequilibrium work
  relation for free energy differences.
\newblock \emph{Physical Review E}, 60\penalty0 (3):\penalty0 2721--2726,
  September 1999.
\newblock ISSN 1063-651X.
\newblock \doi{10.1103/physreve.60.2721}.
\newblock URL \url{http://dx.doi.org/10.1103/physreve.60.2721}.

\bibitem[Seifert(2005)]{Seifert05}
Udo Seifert.
\newblock Entropy production along a stochastic trajectory and an integral
  fluctuation theorem.
\newblock \emph{Phys. Rev. Lett.}, 95\penalty0 (4):\penalty0 040602+, July
  2005.
\newblock \doi{10.1103/physrevlett.95.040602}.
\newblock URL \url{http://dx.doi.org/10.1103/physrevlett.95.040602}.

\bibitem[Hekking and Pekola(2013)]{Hekking13}
F.~W.~J. Hekking and J.~P. Pekola.
\newblock Quantum jump approach for work and dissipation in a two-level system.
\newblock \emph{Phys. Rev. Lett.}, 111:\penalty0 093602+, August 2013.
\newblock \doi{10.1103/physrevlett.111.093602}.
\newblock URL \url{http://dx.doi.org/10.1103/physrevlett.111.093602}.

\bibitem[Alonso et~al.(2016)Alonso, Lutz, and Romito]{Alonso16}
Jose~J. Alonso, Eric Lutz, and Alessandro Romito.
\newblock Thermodynamics of weakly measured quantum systems.
\newblock \emph{Phys. Rev. Lett.}, 116:\penalty0 080403+, February 2016.
\newblock \doi{10.1103/physrevlett.116.080403}.
\newblock URL \url{http://dx.doi.org/10.1103/physrevlett.116.080403}.

\bibitem[Manzano et~al.(2015{\natexlab{a}})Manzano, Horowitz, and
  Parrondo]{Manzano-fluctuation-quantum-maps}
Gonzalo Manzano, Jordan~M. Horowitz, and Juan M~R Parrondo.
\newblock {Nonequilibrium potential and fluctuation theorems for quantum maps}.
\newblock \emph{Physical Review E}, 92\penalty0 (3):\penalty0 032129, sep
  2015{\natexlab{a}}.
\newblock ISSN 1539-3755.
\newblock \doi{10.1103/PhysRevE.92.032129}.
\newblock URL \url{http://arxiv.org/abs/1505.04201
  http://dx.doi.org/10.1103/PhysRevE.92.032129
  https://link.aps.org/doi/10.1103/PhysRevE.92.032129}.

\bibitem[Elouard et~al.(2017{\natexlab{a}})Elouard, Herrera-Mart{\'{i}},
  Clusel, and Auff{\`{e}}ves]{Alexia-measurement-thermodynamics}
Cyril Elouard, David~A. Herrera-Mart{\'{i}}, Maxime Clusel, and Alexia
  Auff{\`{e}}ves.
\newblock {The role of quantum measurement in stochastic thermodynamics}.
\newblock \emph{npj Quantum Information}, 3\penalty0 (1):\penalty0 9, dec
  2017{\natexlab{a}}.
\newblock ISSN 2056-6387.
\newblock \doi{10.1038/s41534-017-0008-4}.
\newblock URL \url{http://arxiv.org/abs/1607.02404
  http://dx.doi.org/10.1038/s41534-017-0008-4
  http://www.nature.com/articles/s41534-017-0008-4}.

\bibitem[Ghe(2017)]{Gherardini17}
Reconstruction of the stochastic quantum entropy production to probe
  irreversibility and correlations, June 2017.
\newblock URL \url{http://arxiv.org/abs/1706.02193}.

\bibitem[Dressel et~al.(2017)Dressel, Chantasri, Jordan, and
  Korotkov]{Dressel17}
Justin Dressel, Areeya Chantasri, Andrew~N. Jordan, and Alexander~N. Korotkov.
\newblock Arrow of time for continuous quantum measurement.
\newblock \emph{Physical Review Letters}, 119\penalty0 (22), December 2017.
\newblock \doi{10.1103/physrevlett.119.220507}.
\newblock URL \url{http://dx.doi.org/10.1103/physrevlett.119.220507}.

\bibitem[Manikandan and Jordan(2018)]{Manikandan18}
Sreenath~K. Manikandan and Andrew~N. Jordan.
\newblock Time reversal symmetry of generalized quantum measurements with past
  and future boundary conditions, January 2018.
\newblock URL \url{http://arxiv.org/abs/1801.04364}.

\bibitem[{Breuer} and {Petruccione}(2007)]{breuer}
Heinz-Peter {Breuer} and Francesco {Petruccione}.
\newblock \emph{{T}he {T}heory of {O}pen {Q}uantum {S}ystems}.
\newblock Oxford University Press, 2007.

\bibitem[Heinosaari and Ziman(2011)]{Heinosaari}
Teiko Heinosaari and Mario Ziman.
\newblock \emph{{T}he {M}athematical {L}anguage of {Q}uantum {T}heory}.
\newblock Cambridge University Press, 2011.

\bibitem[{Stinespring}(1955)]{Stinespring}
W.~F. {Stinespring}.
\newblock Positive functions on $c^*$-algebras.
\newblock \emph{Proc. Amer. Math. Soc.}, 6(2):\penalty0 211--216, 1955.

\bibitem[{Kraus}(1983)]{Kraus}
K.~{Kraus}.
\newblock \emph{{S}tates, {E}ffects and {O}perations: {F}undamental {N}otions
  of {Q}uantum {T}heory}.
\newblock Springer Verlag, 1983.

\bibitem[{Busch} et~al.(1995){Busch}, {Grabowski}, and
  {Lahti}]{Busch-operational}
Paul {Busch}, Marian {Grabowski}, and Pekka~J. {Lahti}.
\newblock \emph{{O}perational {Q}uantum {P}hysics}.
\newblock Springer, 1995.

\bibitem[{Busch} et~al.(2016){Busch}, {Lahti}, {Pellonp\"a\"a}, and
  {Ylinen}]{Busch-measurement-2}
Paul {Busch}, Pekka~J. {Lahti}, Juha~Pekka {Pellonp\"a\"a}, and Kari {Ylinen}.
\newblock \emph{{Q}uantum {M}easurement}.
\newblock Springer, 2016.

\bibitem[Wiseman and Milburn(2010)]{WisemanBook}
H.~M. Wiseman and G.~J. Milburn.
\newblock \emph{Quantum measurement and control}.
\newblock Cambridge University Press, 2010.
\newblock ISBN 9780521804424.
\newblock URL \url{http://www.worldcat.org/isbn/9780521804424}.

\bibitem[Jacobs and Steck(2006)]{Steck06}
Kurt Jacobs and Daniel~A. Steck.
\newblock A straightforward introduction to continuous quantum measurement.
\newblock \emph{Contemporary Physics}, 47\penalty0 (5):\penalty0 279--303,
  September 2006.
\newblock ISSN 0010-7514.
\newblock \doi{10.1080/00107510601101934}.
\newblock URL \url{http://dx.doi.org/10.1080/00107510601101934}.

\bibitem[Cohen-Tannoudji et~al.(1998)Cohen-Tannoudji, Dupont-Roc, and
  Grynberg]{cohen}
Claude Cohen-Tannoudji, Jacques Dupont-Roc, and Gilbert Grynberg.
\newblock \emph{Atom—Photon Interactions : Basic Process and Applications}.
\newblock Wiley-VCH Verlag GmbH, Weinheim, Germany, April 1998.
\newblock ISBN 9783527617197.
\newblock \doi{10.1002/9783527617197}.
\newblock URL \url{http://dx.doi.org/10.1002/9783527617197}.

\bibitem[Lindblad(1976)]{Lindblad76}
G.~Lindblad.
\newblock On the generators of quantum dynamical semigroups.
\newblock \emph{Communications in Mathematical Physics}, 48\penalty0
  (2):\penalty0 119--130, June 1976.
\newblock ISSN 0010-3616.
\newblock \doi{10.1007/bf01608499}.
\newblock URL \url{http://dx.doi.org/10.1007/bf01608499}.

\bibitem[Haroche and Raimond(2006)]{Haroche}
S.~Haroche and J.~M. Raimond.
\newblock \emph{Exploring the quantum : atoms, cavities and photons}.
\newblock Oxford University Press, 2006.
\newblock ISBN 9780198509141.
\newblock URL \url{http://www.worldcat.org/isbn/9780198509141}.

\bibitem[Kist et~al.(1999)Kist, Orszag, Brun, and Davidovich]{Kist99}
Tarso B.~L. Kist, M.~Orszag, T.~A. Brun, and L.~Davidovich.
\newblock Stochastic schr\"{o}dinger equations in cavity qed: physical
  interpretation and localization.
\newblock \emph{Journal of Optics B: Quantum and Semiclassical Optics},
  1\penalty0 (2):\penalty0 251--263, April 1999.
\newblock ISSN 1464-4266.
\newblock \doi{10.1088/1464-4266/1/2/009}.
\newblock URL \url{http://dx.doi.org/10.1088/1464-4266/1/2/009}.

\bibitem[Brun(2000)]{Brun00}
Todd~A. Brun.
\newblock Continuous measurements, quantum trajectories, and decoherent
  histories.
\newblock \emph{Physical Review A}, 61\penalty0 (4), March 2000.
\newblock ISSN 1050-2947.
\newblock \doi{10.1103/physreva.61.042107}.
\newblock URL \url{http://dx.doi.org/10.1103/physreva.61.042107}.

\bibitem[Gardiner and Zoller(2004)]{Gardiner}
Crispin~W. Gardiner and Peter Zoller.
\newblock \emph{Quantum Noise}.
\newblock Springer Berlin Heidelberg, Berlin, Heidelberg, 2004.
\newblock ISBN 978-3-662-04105-5.
\newblock \doi{10.1007/978-3-662-04103-1}.
\newblock URL \url{http://dx.doi.org/10.1007/978-3-662-04103-1}.

\bibitem[Gherardini et~al.(2018)Gherardini, Batalhao, and Paternostro]{CH15}
Stephano Gherardini, Tiago Batalhao, and Mauro Paternostro.
\newblock Measuring irreversibility in open quantum system, in {\it
  `thermodynamics in the quantum regime: recent progress and outlook'},
  springer, 2018.

\bibitem[Crooks(2008)]{Crooks08}
Gavin~E. Crooks.
\newblock Quantum operation time reversal.
\newblock \emph{Phys. Rev. A}, 77:\penalty0 034101+, March 2008.
\newblock \doi{10.1103/physreva.77.034101}.
\newblock URL \url{http://dx.doi.org/10.1103/physreva.77.034101}.

\bibitem[Manzano et~al.(2015{\natexlab{b}})Manzano, Horowitz, and
  Parrondo]{Manzano15}
Gonzalo Manzano, Jordan~M. Horowitz, and Juan M.~R. Parrondo.
\newblock Nonequilibrium potential and fluctuation theorems for quantum maps.
\newblock \emph{Physical Review E}, 92\penalty0 (3), September
  2015{\natexlab{b}}.
\newblock ISSN 1539-3755.
\newblock \doi{10.1103/physreve.92.032129}.
\newblock URL \url{http://dx.doi.org/10.1103/physreve.92.032129}.

\bibitem[Barra and Lled\'{o}(2017)]{Barra17}
Felipe Barra and Crist\'{o}bal Lled\'{o}.
\newblock Stochastic thermodynamics of quantum maps with and without
  equilibrium.
\newblock \emph{Physical Review E}, 96\penalty0 (5), November 2017.
\newblock ISSN 2470-0045.
\newblock \doi{10.1103/physreve.96.052114}.
\newblock URL \url{http://dx.doi.org/10.1103/physreve.96.052114}.

\bibitem[Horowitz(2012)]{Horowitz12}
Jordan~M. Horowitz.
\newblock Quantum-trajectory approach to the stochastic thermodynamics of a
  forced harmonic oscillator.
\newblock \emph{Phys. Rev. E}, 85:\penalty0 031110+, March 2012.
\newblock \doi{10.1103/physreve.85.031110}.
\newblock URL \url{http://dx.doi.org/10.1103/physreve.85.031110}.

\bibitem[Manzano et~al.(2017)Manzano, Horowitz, and Parrondo]{Manzano17}
Gonzalo Manzano, Jordan~M. Horowitz, and Juan M.~R. Parrondo.
\newblock Quantum fluctuation theorems for arbitrary environments: adiabatic
  and non-adiabatic entropy production, September 2017.
\newblock URL \url{http://arxiv.org/abs/1710.00054}.

\bibitem[{P}etz(2008)]{Petz-QI}
D\'{e}nes {P}etz.
\newblock \emph{{Q}uantum {I}nformation {T}heory and {Q}uantum {S}tatistics}.
\newblock Springer, 2008.

\bibitem[Alberti and Uhlmann(1982)]{Uhlmann-Stochasticity}
P.M. Alberti and A.~Uhlmann.
\newblock \emph{{S}tochasticity and {P}artial {O}rder: {D}oubly {S}tochastic
  {M}aps and {U}nitary {M}ixing}.
\newblock Springer, 1982.

\bibitem[Nakahara et~al.(2008)Nakahara, Rahimi, and
  Saitoh]{Nakahara-Decoherence}
M.~Nakahara, R.~Rahimi, and A.~Saitoh.
\newblock \emph{{D}ecoherence {S}uppression in {Q}uantum {S}ystems}.
\newblock World Scientific, 2008.

\bibitem[Uzdin(2018)]{CH27}
Raam Uzdin.
\newblock The second law and beyond in microscopic quantum setups, in {\it
  `thermodynamics in the quantum regime: recent progress and outlook'}, 2018.

\bibitem[Murashita(2015)]{MurashitaThesis}
Y\^{u}to Murashita.
\newblock Absolute irreversibility in information thermodynamics.
\newblock Master's thesis, June 2015.
\newblock URL \url{http://arxiv.org/abs/1506.04470}.

\bibitem[Funo et~al.(2015)Funo, Murashita, and Ueda]{Funo15}
Ken Funo, Y\^{u}to Murashita, and Masahito Ueda.
\newblock Quantum nonequilibrium equalities with absolute irreversibility.
\newblock \emph{New Journal of Physics}, 17\penalty0 (7):\penalty0 075005+,
  July 2015.
\newblock ISSN 1367-2630.
\newblock \doi{10.1088/1367-2630/17/7/075005}.
\newblock URL \url{http://dx.doi.org/10.1088/1367-2630/17/7/075005}.

\bibitem[Tas(2000)]{Tasaki00}
Jarzynski relations for quantum systems and some applications, September 2000.
\newblock URL \url{http://arxiv.org/abs/cond-mat/0009244.pdf}.

\bibitem[Kurchan(2001)]{Kurchan01}
Jorge Kurchan.
\newblock A quantum fluctuation theorem, August 2001.
\newblock URL \url{http://arxiv.org/abs/cond-mat/0007360}.

\bibitem[Mukamel(2003)]{Mukamel03}
Shaul Mukamel.
\newblock Quantum extension of the jarzynski relation: Analogy with stochastic
  dephasing.
\newblock \emph{Physical Review Letters}, 90\penalty0 (17), May 2003.
\newblock \doi{10.1103/physrevlett.90.170604}.
\newblock URL \url{http://dx.doi.org/10.1103/physrevlett.90.170604}.

\bibitem[Talkner et~al.(2007)Talkner, Lutz, and H\"{a}nggi]{Talkner07}
Peter Talkner, Eric Lutz, and Peter H\"{a}nggi.
\newblock Fluctuation theorems: Work is not an observable.
\newblock \emph{Physical Review E}, 75\penalty0 (5):\penalty0 050102+, May
  2007.
\newblock ISSN 1539-3755.
\newblock \doi{10.1103/physreve.75.050102}.
\newblock URL \url{http://dx.doi.org/10.1103/physreve.75.050102}.

\bibitem[Campisi et~al.(2011)Campisi, H\"{a}nggi, and Talkner]{Campisi11}
Michele Campisi, Peter H\"{a}nggi, and Peter Talkner.
\newblock Colloquium : Quantum fluctuation relations: Foundations and
  applications.
\newblock \emph{Reviews of Modern Physics}, 83\penalty0 (3):\penalty0 771--791,
  July 2011.
\newblock ISSN 0034-6861.
\newblock \doi{10.1103/revmodphys.83.771}.
\newblock URL \url{http://dx.doi.org/10.1103/revmodphys.83.771}.

\bibitem[Funo et~al.(2018)Funo, Sagawa, and Ueda]{CH9}
Ken Funo, Takahiro Sagawa, and Masahito Ueda.
\newblock Quantum fluctuation theorems, in {\it `thermodynamics in the quantum
  regime: recent progress and outlook'}, 2018.

\bibitem[Alicki and Kosloff(2018)]{intro}
Robert Alicki and Ronnie Kosloff.
\newblock Introduction to quantum thermodynamics: History and prospects, in
  {\it `thermodynamics in the quantum regime: recent progress and outlook'},
  2018.

\bibitem[Esposito et~al.(2009)Esposito, Harbola, and Mukamel]{Esposito09}
Massimiliano Esposito, Upendra Harbola, and Shaul Mukamel.
\newblock Nonequilibrium fluctuations, fluctuation theorems, and counting
  statistics in quantum systems.
\newblock \emph{Rev. Mod. Phys.}, 81\penalty0 (4):\penalty0 1665--1702,
  December 2009.
\newblock \doi{10.1103/revmodphys.81.1665}.
\newblock URL \url{http://dx.doi.org/10.1103/revmodphys.81.1665}.

\bibitem[Elouard et~al.(2018)Elouard, Hererra-Mart\'i, and
  Auff\`eves]{Elouard18Fluo}
Cyril Elouard, David Hererra-Mart\'i, and Alexia Auff\`eves.
\newblock Thermodynamics of fluorescence, in preparation, 2018.

\bibitem[Elouard(2017)]{ElouardThesis}
Cyril Elouard.
\newblock Thermodynamics of quantum open systems: Applications in quantum
  optics and optomechanics, September 2017.
\newblock URL \url{http://arxiv.org/abs/1709.02744}.

\bibitem[Elouard et~al.(2017{\natexlab{b}})Elouard, Herrera-Mart{\'{i}}, Huard,
  and Auff{\`{e}}ves]{Elouard2017a}
Cyril Elouard, David Herrera-Mart{\'{i}}, Benjamin Huard, and Alexia
  Auff{\`{e}}ves.
\newblock {Extracting Work from Quantum Measurement in Maxwell's Demon
  Engines}.
\newblock \emph{Physical Review Letters}, 118\penalty0 (26):\penalty0 260603,
  jun 2017{\natexlab{b}}.
\newblock ISSN 0031-9007.
\newblock \doi{10.1103/PhysRevLett.118.260603}.
\newblock URL \url{http://arxiv.org/abs/1702.01917
  http://link.aps.org/doi/10.1103/PhysRevLett.118.260603}.

\bibitem[Yi et~al.(2017)Yi, Talkner, and Kim]{Yi17}
Juyeon Yi, Peter Talkner, and Yong~W. Kim.
\newblock Single-temperature quantum engine without feedback control.
\newblock \emph{Physical Review E}, 96\penalty0 (2), August 2017.
\newblock ISSN 2470-0045.
\newblock \doi{10.1103/physreve.96.022108}.
\newblock URL \url{http://dx.doi.org/10.1103/physreve.96.022108}.

\bibitem[Elouard and Jordan(2018)]{Elouard18}
Cyril Elouard and Andrew~N. Jordan.
\newblock Efficient quantum measurement engine, January 2018.
\newblock URL \url{http://arxiv.org/abs/1801.03979}.

\bibitem[Elouard et~al.(2017{\natexlab{c}})Elouard, Bernardes, Carvalho,
  Santos, and Auff{\`{e}}ves]{Alexia-fluctuation-engineered-reservoir}
C.~Elouard, N.~K. Bernardes, A.~R.~R. Carvalho, M.~F. Santos, and
  A.~Auff{\`{e}}ves.
\newblock {Probing quantum fluctuation theorems in engineered reservoirs}.
\newblock \emph{New Journal of Physics}, 19\penalty0 (10):\penalty0 103011, oct
  2017{\natexlab{c}}.
\newblock ISSN 1367-2630.
\newblock \doi{10.1088/1367-2630/aa7fa2}.
\newblock URL \url{http://arxiv.org/abs/1702.06811
  http://dx.doi.org/10.1088/1367-2630/aa7fa2
  http://stacks.iop.org/1367-2630/19/i=10/a=103011?key=crossref.c896ebe28f61f365f21ff0014939a2f7}.

\bibitem[Ghe(2018)]{Gherardini18}
Non-equilibrium quantum-heat statistics under stochastic projective
  measurements, May 2018.
\newblock URL \url{http://arxiv.org/abs/1805.00773}.

\bibitem[Mohammady et~al.(2018)Mohammady, Auff\'eves, and Anders]{Mohammady18}
Hamed Mohammady, Alexia Auff\'eves, and Janet Anders.
\newblock in preparation, 2018.

\bibitem[Pekola et~al.(2013)Pekola, Solinas, Shnirman, and Averin]{Pekola13}
J.~P. Pekola, P.~Solinas, A.~Shnirman, and D.~V. Averin.
\newblock Calorimetric measurement of work in a quantum system.
\newblock \emph{New Journal of Physics}, 15\penalty0 (11):\penalty0 115006+,
  November 2013.
\newblock ISSN 1367-2630.
\newblock \doi{10.1088/1367-2630/15/11/115006}.
\newblock URL \url{http://dx.doi.org/10.1088/1367-2630/15/11/115006}.

\bibitem[Szczygielski et~al.(2013)Szczygielski, Klimovsky, and
  Alicki]{Alicki13}
Krzysztof Szczygielski, David~G. Klimovsky, and Robert Alicki.
\newblock Markovian master equation and thermodynamics of a two-level system in
  a strong laser field.
\newblock \emph{Phys. Rev. E}, 87:\penalty0 012120+, January 2013.
\newblock \doi{10.1103/physreve.87.012120}.
\newblock URL \url{http://dx.doi.org/10.1103/physreve.87.012120}.

\bibitem[Cuetara et~al.(2015)Cuetara, Engel, and Esposito]{Bulnes15}
Gregory~B. Cuetara, Andreas Engel, and Massimiliano Esposito.
\newblock Stochastic thermodynamics of rapidly driven systems.
\newblock \emph{New Journal of Physics}, 17\penalty0 (5):\penalty0 055002+, May
  2015.
\newblock ISSN 1367-2630.
\newblock \doi{10.1088/1367-2630/17/5/055002}.
\newblock URL \url{http://dx.doi.org/10.1088/1367-2630/17/5/055002}.

\bibitem[Donvil(2018)]{Donvil18}
Brecht Donvil.
\newblock Thermodynamics of a periodically driven qubit.
\newblock \emph{Journal of Statistical Mechanics: Theory and Experiment},
  2018\penalty0 (4):\penalty0 043104+, April 2018.
\newblock ISSN 1742-5468.
\newblock \doi{10.1088/1742-5468/aab857}.
\newblock URL \url{http://dx.doi.org/10.1088/1742-5468/aab857}.

\bibitem[Naghiloo et~al.(2017)Naghiloo, Tan, Harrington, Alonso, Lutz, Romito,
  and Murch]{Naghiloo17}
M.~Naghiloo, D.~Tan, P.~M. Harrington, J.~J. Alonso, E.~Lutz, A.~Romito, and
  K.~W. Murch.
\newblock Thermodynamics along individual trajectories of a quantum bit,
  September 2017.
\newblock URL \url{http://arxiv.org/abs/1703.05885}.

\bibitem[Campisi et~al.(2009)Campisi, Talkner, and H\"{a}nggi]{Campisi09}
Michele Campisi, Peter Talkner, and Peter H\"{a}nggi.
\newblock Fluctuation theorem for arbitrary open quantum systems.
\newblock \emph{Physical Review Letters}, 102\penalty0 (21):\penalty0 210401+,
  May 2009.
\newblock ISSN 0031-9007.
\newblock \doi{10.1103/physrevlett.102.210401}.
\newblock URL \url{http://dx.doi.org/10.1103/physrevlett.102.210401}.

\bibitem[Didier et~al.(2015)Didier, Bourassa, and Blais]{Didier15}
Nicolas Didier, J\'{e}r\^{o}me Bourassa, and Alexandre Blais.
\newblock Fast quantum nondemolition readout by parametric modulation of
  longitudinal qubit-oscillator interaction.
\newblock \emph{Phys. Rev. Lett.}, 115\penalty0 (20):\penalty0 203601+,
  November 2015.
\newblock ISSN 0031-9007.
\newblock \doi{10.1103/physrevlett.115.203601}.
\newblock URL \url{http://dx.doi.org/10.1103/physrevlett.115.203601}.

\bibitem[Naghiloo et~al.(2018)Naghiloo, Alonso, Romito, Lutz, and
  Murch]{Naghiloo18}
M.~Naghiloo, J.~J. Alonso, A.~Romito, E.~Lutz, and K.~W. Murch.
\newblock Information gain and loss for a quantum maxwell's demon, February
  2018.
\newblock URL \url{http://arxiv.org/abs/1802.07205}.

\bibitem[Jordan et~al.(2016)Jordan, Chantasri, Rouchon, and Huard]{Chantasri16}
AndrewN Jordan, Areeya Chantasri, Pierre Rouchon, and Benjamin Huard.
\newblock Anatomy of fluorescence: quantum trajectory statistics from
  continuously measuring spontaneous emission.
\newblock 3\penalty0 (3):\penalty0 237--263, 2016.
\newblock \doi{10.1007/s40509-016-0075-9}.
\newblock URL \url{http://dx.doi.org/10.1007/s40509-016-0075-9}.

\bibitem[Garner(2018)]{CH13}
Andrew Garner.
\newblock One-shot information-theoretic approaches to fluctuation theorems, in
  {\it `thermodynamics in the quantum regime: recent progress and outlook'},
  2018.

\bibitem[W\"{a}chtler et~al.(2016)W\"{a}chtler, Strasberg, and
  Brandes]{Wachtler16}
Christopher~W. W\"{a}chtler, Philipp Strasberg, and Tobias Brandes.
\newblock Stochastic thermodynamics based on incomplete information:
  generalized jarzynski equality with measurement errors with or without
  feedback.
\newblock \emph{New Journal of Physics}, 18\penalty0 (11):\penalty0 113042+,
  November 2016.
\newblock ISSN 1367-2630.
\newblock \doi{10.1088/1367-2630/18/11/113042}.
\newblock URL \url{http://dx.doi.org/10.1088/1367-2630/18/11/113042}.

\bibitem[Di\'{o}si and Strunz(1997)]{Diosi97}
Lajos Di\'{o}si and Walter~T. Strunz.
\newblock The non-markovian stochastic schr\"{o}dinger equation for open
  systems.
\newblock \emph{Physics Letters A}, 235\penalty0 (6):\penalty0 569--573,
  November 1997.
\newblock ISSN 03759601.
\newblock \doi{10.1016/s0375-9601(97)00717-2}.
\newblock URL \url{http://dx.doi.org/10.1016/s0375-9601(97)00717-2}.

\bibitem[Strunz et~al.(1999)Strunz, Di\'{o}si, and Gisin]{Strunz99}
Walter~T. Strunz, Lajos Di\'{o}si, and Nicolas Gisin.
\newblock Open system dynamics with non-markovian quantum trajectories.
\newblock \emph{Physical Review Letters}, 82\penalty0 (9):\penalty0 1801--1805,
  March 1999.
\newblock ISSN 0031-9007.
\newblock \doi{10.1103/physrevlett.82.1801}.
\newblock URL \url{http://dx.doi.org/10.1103/physrevlett.82.1801}.

\bibitem[Jack and Collett(2000)]{Jack00}
M.~W. Jack and M.~J. Collett.
\newblock Continuous measurement and non-markovian quantum trajectories.
\newblock \emph{Physical Review A}, 61\penalty0 (6), May 2000.
\newblock ISSN 1050-2947.
\newblock \doi{10.1103/physreva.61.062106}.
\newblock URL \url{http://dx.doi.org/10.1103/physreva.61.062106}.

\bibitem[Jing et~al.(2013)Jing, Zhao, You, Strunz, and Yu]{Jing13}
Jun Jing, Xinyu Zhao, J.~Q. You, Walter~T. Strunz, and Ting Yu.
\newblock Many-body quantum trajectories of non-markovian open systems.
\newblock \emph{Physical Review A}, 88\penalty0 (5), November 2013.
\newblock ISSN 1050-2947.
\newblock \doi{10.1103/physreva.88.052122}.
\newblock URL \url{http://dx.doi.org/10.1103/physreva.88.052122}.

\end{thebibliography}

\section*{Appendices}

\subsection*{A. Trajectories in presence of non rank-1 measurements}

In Section \ref{s:Qtraj}, we have restricted our discussion to the case where both system and environment are measured with rank-1 projective measurements, which allowed us to obtain pure-state trajectories for both system and environment. While this describes very well certain measurement schemes like an heterodyne setup \cite{Chantasri16}, such measurements are unfeasible in many physical situations which still allow to track quantum trajectories. Especially, this is the case when the environment is a thermal bath with infinite degrees of freedom. To model such cases, we have to allow not all the different pairs of outcomes $(\mu,\nu)$ to be distinguished by the measurement. We define a set of ``macroscopic'' outcomes $\{\alpha\}$ that can indeed by deduced from the measurement, where each $\alpha$ contains several pairs $(\mu,\nu)$. This leads to the decomposition of the quantum map $\Phi$ as $\Phi(\rho\sub{\s}) = \sum_{\alpha} \Phi_\alpha(\rho\sub{\s})$, where
\bb\label{eq:Phialpha}
\Phi_\alpha(\rho\sub{\s}) = \sum_{(\mu,\nu) \in \alpha } \Phi_{\mu,\nu}(\rho\sub{\s}).
\ee
Such ``coarse-graining'' prevents us in general to write $\Phi_\alpha$ under the form \eqref{eq:Phimunu}. Indeed, $\Phi_\alpha$ will in general be described by a set of non-identical Kraus operators. As such, even if the system is initially in a pure state $\prs{\psi}$, $\Phi_\alpha(\prs{\psi})$ will in general be mixed. Remarkably however, some practical measurement schemes (e.g. the quantum jump unraveling) still allow us to define Kraus operators $M_\alpha$ fulfilling $\Phi_\alpha(\rho\sub{\s}) = M_\alpha \rho\sub{\s} M_\alpha^\dagger$. 
Of particular interest are measurement schemes such that for every $(\mu,\nu)$ and $(\mu', \nu')$ in $\alpha$, $M_{\mu',\nu'} = \sqrt{\tr[M_{\mu',\nu'}^\dagger M_{\mu',\nu'}]/\tr[M_{\mu,\nu}^\dagger M_{\mu,\nu}]} M_{\mu,\nu}$, allowing us to define
\bb\label{eq:Malpha}
M_\alpha := \frac{\sqrt{q_\alpha} M_{\mu,\nu}}{\sqrt{\tr[M_{\mu,\nu}^\dagger M_{\mu,\nu}]}}
\ee
for any $(\mu, \nu)$ in $\alpha$,  where $q_\alpha := \sum_{(\mu,\nu)\in\alpha} \tr[M_{\mu,\nu}^\dagger M_{\mu,\nu}]$. It is then easy to verify that $M_\alpha \rho\sub{\s} M_\alpha^\dagger = \sum_{(\mu,\nu) \in \alpha}M_{\mu,\nu} \rho\sub{\s} M_{\mu,\nu}^\dagger$. This ensures that such a coarse-graining  of the environment trajectories will not alter the average evolution of the system, which will still be described by the quantum channel $\Phi$.

The case of a quantum jump unraveling of the Lindblad equation can be modeled by considering the macroscopic outcomes are the variations of the number of excitations in the environment between $\Delta t$. One only needs to know the variation $\alpha$ and not the absolute number of excitations, neither at the beginning nor at the end of the time step to characterize the system's evolution. Consequently, the Kraus operators $M_\alpha$ are composed of several of the operators $M_{\mu,\nu}$ that would be obtained by taking $\{\phi_\mu\}$ and $\{\varphi_\nu\}$ to be the number of excitations basis. With such a definition of $M_{\mu,\nu}$, the QJ Kraus operators obey Eq.\eqref{eq:Malpha}. As every $M_{\mu,\nu}$ for $(\mu,\nu)\in\alpha$ is proportional to the same jump operator $L_j$, we may still view the system's evolution as a trajectory of pure states.

\subsection*{B. Entropy production in presence of non rank-1 measurements}\label{s:TRalpha}

The results of section \ref{s:reversed-traj-entropy-prod} can be extended to the case discussed in Appendix A, wherein the measurement outcomes are coarse-grained so that $\tilde \Phi_{\alpha_k} = \sum_{(\mu_k,\nu_k)\in{\alpha_k}} \tilde\Phi_{\mu_k,\nu_k}$. In the case where the Kraus operators $M_{\alpha_k}$ can be defined by Eq.\eqref{eq:Malpha} (e.g. the QJ unraveling), the Kraus operator for the reverse transformation will be
\begin{align}
\tilde M_{\alpha_k} = \sqrt{\tilde q_{\alpha_k}}\frac{ M_{\mu_k,\nu_k}^\dagger}{ \sqrt{\tr[ M_{\mu_k,\nu_k}^\dagger M_{\mu_k,\nu_k}]}}
\end{align}
for any $(\mu_k, \nu_k) \in \alpha_k$, where we have introduced 
\begin{align}
\tilde q_{\alpha_k} &:= \sum_{(\mu_k,\nu_k)\in{\alpha_k}} \tr[\tilde M_{\mu_k,\nu_k}^\dagger \tilde M_{\mu_k,\nu_k}], \nonumber \\
&= \sum_{(\mu_k,\nu_k)\in{\alpha_k}} \frac{q_{\nu_k}'}{q_{\mu_k}}\tr[ M_{\mu_k,\nu_k}^\dagger  M_{\mu_k,\nu_k}].
\end{align} 

Consequently, the expression for stochastic entropy production along the trajectories $\Gamma = ((l,m), \vec \alpha)$ will be 
\begin{align}
\Delta_i s[\Gamma] &=  \ln{\frac{p_l}{p_m'}} + \sum_{k=1}^K \ln{\frac{q_{\alpha_k}}{\tilde q_{\alpha_k}}}.
\end{align}
Note that while the entropy production contribution from the system depends on its probability distribution $p_l$ and $p_m'$, at the start and end of the transformation, this is not so for the bath contributions. Notwithstanding,  the IFT Eq.\eqref{eq:IFT} is still valid, while the average entropy production now reads:
\begin{align}
\label{eq:DiSalpha}
\moy{\Delta_i s[\Gamma]} &=  \Delta S\sub{\s} + \sum_{k=1}^K \sum_{\alpha_k} \frac{\tr[M_{\mu,\nu}^\dagger M_{\mu,\nu} \rho\sub{\s}^{(k)} ]}{ \tr[M_{\mu,\nu}^\dagger M_{\mu,\nu}]} q_{\alpha_k}(\ln{q_{\alpha_k}} - \ln{\tilde q_{\alpha_k}}),
\end{align}
where $\rho\sub{\s}^{(k)} := (\Phi(t_{k-1},t_k) \circ \dots \Phi(t_0,t_1) )(\rho\sub{\s})$ is the average state of the system before it interacts with the $k$\ts{th} environment.
In contrast with Eq.\eqref{eq:DiS2.1}, the second term in the right-hand side of Eq.\eqref{eq:DiSalpha} is not equal in general to the variation of the environment's von-Neumann entropy. This is a consequence of the uncertainty introduced by the ``coarse-graining'' present in the measurement scheme, i.e. the fact that not all the couples $(\mu,\nu)$ of outcomes can be discriminated. However, certain measurement schemes like the QJ unraveling for a qubit presented in Section \ref{s:Ex2} actually gather enough information so that the second term in \eqref{eq:DiSalpha} reduces to $\Delta S_\e$.


\end{document}